\documentclass[aps,prl,twocolumn,floats]{revtex4}
\usepackage{latexsym}
\usepackage{amsmath}
\usepackage{amssymb}
\usepackage{bm}
\usepackage{wasysym}
\usepackage[dvips]{color}
\usepackage{graphicx}
\usepackage{graphicx}
\usepackage{subfigure}
\usepackage{epsfig}
\newcommand{\hide}[1]{}
\newcommand{\ua}{\uparrow}
\newcommand{\da}{\downarrow}
\newcommand{\veps}{\varepsilon}
\def\ra{\rangle}
\def\la{\langle}

\def\veps{\varepsilon}

\begin{document}
\title{A Model for Two-Channel Kondo Effect in CNT Quantum Dot}

\author{I. Kuzmenko, T. Kuzmenko and Y. Avishai}
\affiliation{\footnotesize
  Department of Physics, Ben-Gurion University of
  the Negev, Beer-Sheva, Israel 
  }

\date{\today}

\begin{abstract}
Over-screened Kondo effect is feasible in carbon
nanotube quantum dot junction hosting a spin $\tfrac{1}{2}$ atom 
with single $s$-wave valence electron (e.g Au). The idea is to use the two
valleys as two symmetry protected  flavor quantum numbers
$\xi={\bf K}, {\bf K}'$.  Perturbative RG analysis exposes
the finite weak-coupling two-channel fixed point, where
the Kondo temperature is estimated to be around
$0.5\div5$~K. Remarkably, occurrence of two different scaling regimes 
implies a non-monotonic dependence of the conductance as function of temperature. 
\end{abstract}

\maketitle
\noindent
\underline{\it Introduction} 
About three and a half decades ago, it was noticed \cite{NoBla80} that 
when a magnetic impurity of spin 1/2 is over-screened by two identical
conducting electron channels, the many-body physics is characterized by
a non-Fermi-liquid fixed point at temperature $T$ smaller than
the Kondo temperature $T_K$. Experimentally, this phenomenon, referred to as
two-channel Kondo effect (2CKE), is characterized by unusual physics as $T \to 0$, such as
non-zero entropy, divergence of susceptibility and others
\cite{2CKE-prb11,Logan-prb11,Logan-prb10,Giuliano-JPhysCondMat04,Eran-prb03}.
Realizing 2CKE in quantum dot with odd electron occupation
is remarkably elusive due to channel anisotropy emerging from
inter-channel co-tunneling processes. To remedy this instability,
a suppression of inter-channel co-tunneling is attempted
\cite{Potok07} where the interference is suppressed by Coulomb
blockade.

\begin{figure}
\vspace{-0.2in}
\centering\subfigure[]{\includegraphics[width=0.32\textwidth]
{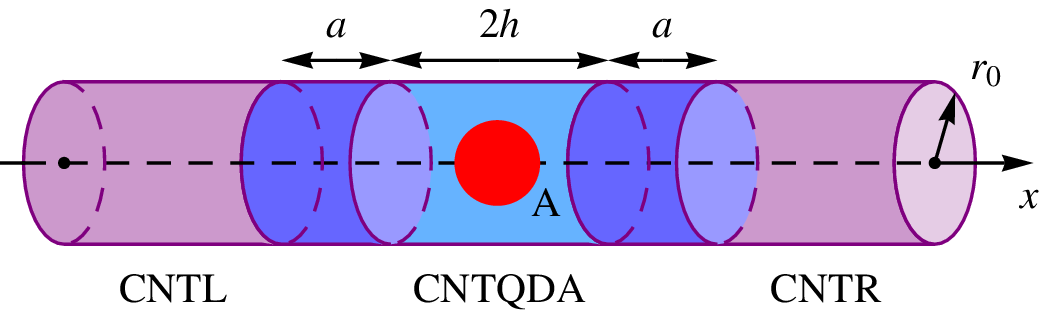}
\label{Fig1a}}
\centering\subfigure[]{\includegraphics[width=0.3\textwidth]
{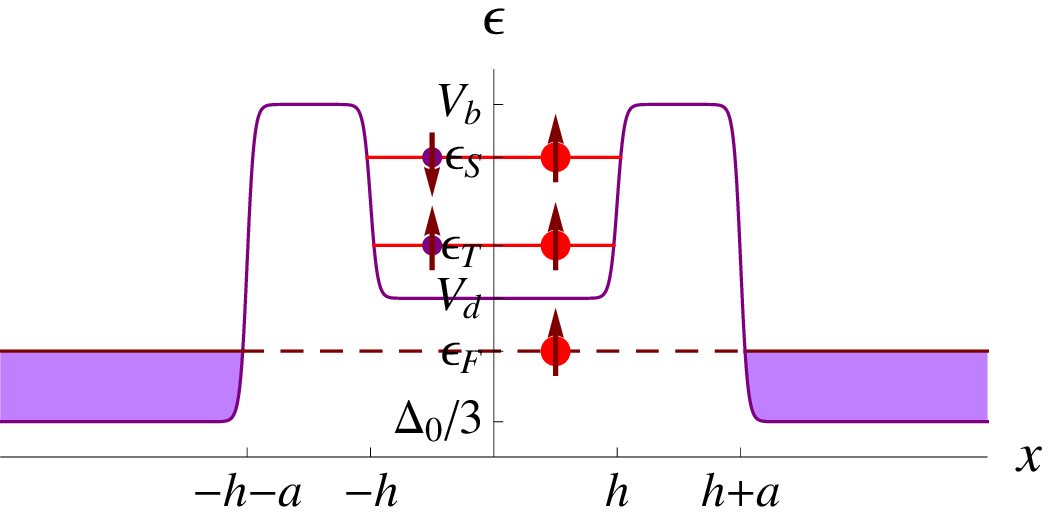}
\label{Fig1b}}
 \caption{\footnotesize
   CNTL-CNTQDA-CNTR junction.
   (a) Schematic geometry of the junction including semi-infinite
   left and right leads, separated from a quantum dot of length
   $2h$ (that hosts a spin 1/2 atom A) by two barriers
   of width $a$.
   (b) Low energy levels of the quantum dot with (from below)
   the caged atom, followed by triplet and singlet atom-electron
   states. }
  \label{Fig1}
\end{figure}

\noindent
\underline{\it In this work} we use a novel approach to avoid
channel mixing using a CNTL-CNTQDA-CNTR  junction as
shown in Fig.~\ref{Fig1a}.
Here CNTL, CNTR are semi infinite carbon nano tube (CNT) left
and right leads and CNTQDA is a short CNT quantum dot with
an atom A with $s$-wave valence  electron of spin $S_A$=$\tfrac{1}{2}$ implanted on its axis.
The two valleys ${\bf K}$ and ${\bf K}'$ serve as two symmetry
protected  flavor quantum numbers $\xi={\bf K}, {\bf K}'$.
The CNTQDA is gated such that its (neutral) ground state 
consists of the caged atom ket $|M_A$=$\pm \tfrac{1}{2} \ra$ and its 
lowest excited (charged)
states are singlet and triplet (defined explicitly below), formed as proper combinations of
basic states  $|M_A\ra\otimes|\xi \sigma\ra$, where $|\xi \sigma\ra$ 
is a CNTQD electron of flavor $\xi$ and spin projection $\sigma=\pm \tfrac{1}{2}$. 
The Anderson model hybridizes lead and dot electrons with
the same flavor and spin projection, and the Schrieffer-Wolf
transformation, while mixing spin projections does not mix flavors,
thereby realizing a two-channel Kondo physics. Perturbative RG
analysis exposes the finite weak-coupling two-channel fixed point,
where the Kondo temperature is estimated to be around
$0.5\div5$~K.\\
It should be stressed that a junction with CNTQD 
containing odd number of electrons in the ground state albeit
{\it without a caged atom},
will {\it not} display the 2CKE, because an electron with
flavor quantum number $\xi$ can tunnel from the dot
to the lead, and be replaced by another electron with flavor
quantum number $\xi'$. As a result, flavor is not
a good quantum number and flavor (channel) mixing results
in crossover to the ordinary (single-channel) 
Kondo effect \cite{GR,KKA02-2chKE}.\\
\noindent
\underline{\it Implanting  Gold Atom on the CNT axis}\\
A crucial ingredient in the present analysis is the feasibility
of caging an atom such as Au (for example) on the CNT
axis $x$ as in Fig.~\ref{Fig1}a. Here we briefly describe the underlying construction. 
Technical aspects of atomic physics are detailed in the Supplemental Material (SM). \\
Consider first an infinitely long CNT (along the $x$ axis) and
let ${\mathbf{q}}={\mathbf k}-{\mathbf K}$ or
${\mathbf{q}}={\mathbf k}-{\mathbf K}'$. For
an electron at small $|{\mathbf{q}}|$, the energy
dispersion is \cite{Ando-93-1,Ando3},
$\epsilon_{qm}=\sqrt{(\hbar v q)^2+(m+\nu)^2\Delta_0^2},$
where ${\mathbf{q}}=(q,\frac{m+\nu}{r_0})$, $q$ is
the wave number in the CNT direction and $m$ is the orbital quantum
number ($m=0,\pm1,\pm2,\ldots$), $\Delta_0=\tfrac{\hbar v}{r_0}$,
$v\sim10^8$~cm/sec is the group
velocity of electrons in a metallic CNT \cite{Ando-93-1,Ando3}
and $r_0$ is the CNT radius.
The quantum number $\nu$ is 0 for metallic CNT or
$\pm\frac{1}{3}$ for semiconductor CNT \cite{Ando-93-1,Ando3}.
In what following, $\nu$ will be tuned to be non-zero,
implying a semiconductor CNT. \\
Now let us check under what conditions it is possible to implant and stabilize 
a gold atom on the CNT axis. 
Denoting the van der Waals interaction between the Au and
C atoms a distance $Y$ apart by $V_{\mathrm{w}}(Y)$,
the van der Waals interaction of the Au atom
with the {\it entire} CNT is then,
\begin{equation}
  V({\mathbf{R}}) =
  \sum_{\alpha=A,B}
  \sum_{{\mathbf{R}}_{\alpha}}
  V_{\mathrm{w}}\big(\big|{\mathbf{R}}-{\mathbf{R}}_{\alpha}\big|\big),
  \label{van-der-Waals-atom-CNT}
\end{equation}
where ${\mathbf{R}}$ is the position of the Au atom,
${\mathbf{R}}_{A,B}$ are the positions of atoms C
on sub-lattices $A,B$. Due to cylindrical symmetry,
$V({\mathbf{R}})=V(R,X)$ depends on $X$ (along which it is periodic)
and $R$ (radial variable).
\begin{figure}[htb]
\centering
\centering\subfigure[]{\includegraphics[width=0.28\textwidth]
{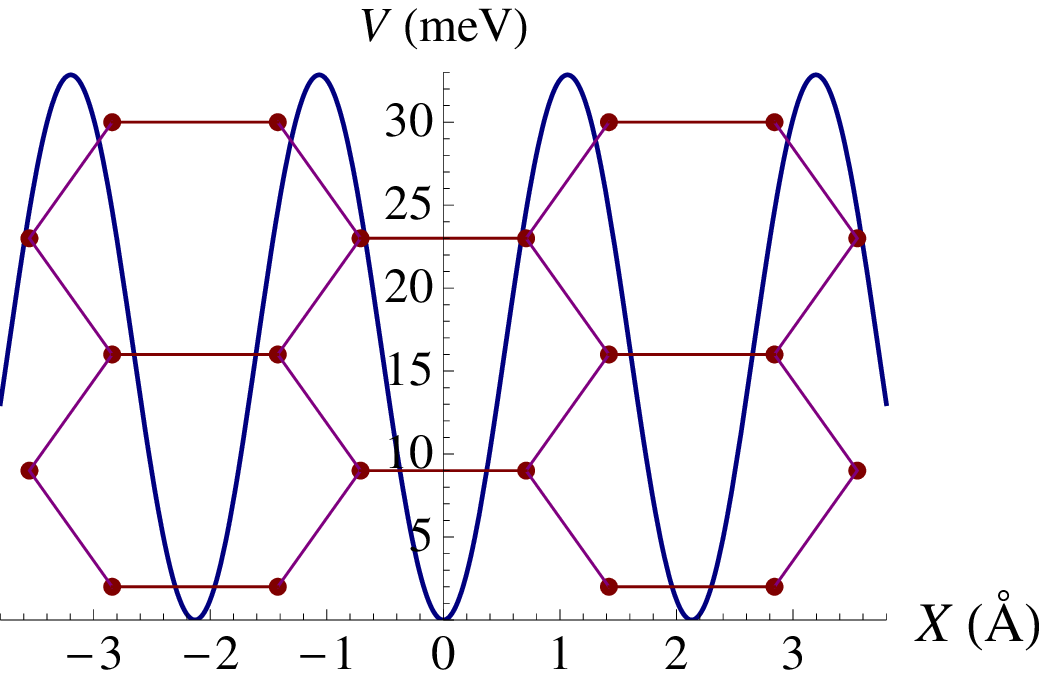}
\label{Fig-vdWa}}
\centering\subfigure[]{\includegraphics[width=0.28\textwidth]
{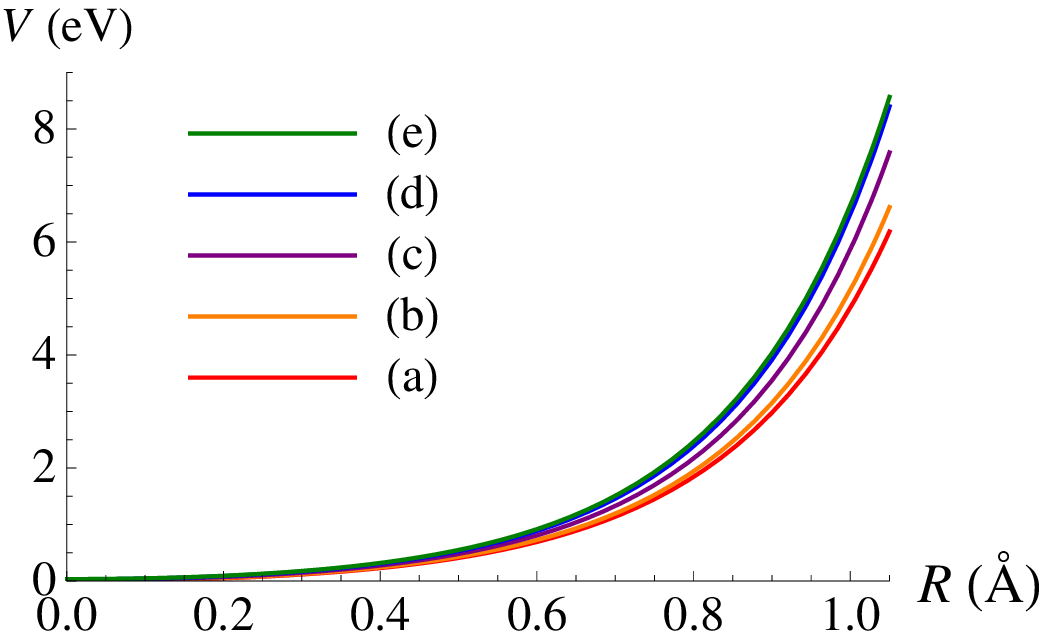}
\label{Fig-vdWb}}
\vspace{-0.1in}
 \caption{\footnotesize
  (Color online)
  (a) Van der Waals potential $V(0,X)$,
  Eq. (\ref{van-der-Waals-atom-CNT}). The purple dots
  and lined denote the lattice of the CNT.
  (b) Van der Waals potential $V(R,X)$ as a function
  of $R$ for different values of $X$: Curves a,b,c,d,e
  correspond to $X=0.25 (m-1)X_1, m=1,2,3,4,5$.}
 \label{Fig-vdW}
\end{figure}
\vspace{-0.1in}
Plots of $V(0,X)$ and $V(R;X)$  [as calculated in the SM
based on realistic parametrization of $V_{\mathrm{w}}(Y)$],
are displayed in Fig. \ref{Fig-vdW}a, b for a zigzag CNT with
radius $r_0=3.13$~{\AA}.
$V(R,X)$ has minima and saddle points at
$(0,X_n^{\mathrm{min}}),  (0,X_n^{\mathrm{sad}})$,
($n=0,\pm 1,\pm 2, \ldots$)
\begin{eqnarray}
  X_n^{\mathrm{min}} =
  2 n X_1,
  \
  X_n^{\mathrm{sad}} ~=~
  (2n+1)X_1,
  \
  X_1=
  \frac{3 a_0}{4\sqrt{3}}. 
  \label{r-min}
\end{eqnarray}
The height of the tunnel barriers between two minima is
\begin{eqnarray}
  W_b &=&V(0,X_n^{\mathrm{sad}})=
  32.8~{\mathrm{meV}}.
  \label{V-barrier-append}
\end{eqnarray}
Expansion of $V(R,X)$ around the minima
$(0,X_n^{\mathrm{min}})$ shows that, to second
order,  the Au atom moves in an anisotropic harmonic
oscillator potential with spring constants
$K_x=0.072$~eV/\AA$^2$, $K_r=1.117$~eV/\AA$^2$.
The energy of the ground state in this harmonic potential
is about $5.5$~meV, much smaller than the barrier height
$W_b$. Moreover, the harmonic radius  $a_x\approx0.11$~\AA,
is much smaller than $X_1=1.07$~{\text{\AA}}, so that
the harmonic approximation is self-consistent and
the picture of well localized Au atom on the CNT axis makes sense.

\noindent
\underline{\it Gating a Quantum Dot with implanted Magnetic Atom:} 
To get a CNTL-CNTQDA-CNTR 
junction as in Fig.~\ref{Fig1a} the infinite CNT is now gated by the potentil:
\begin{eqnarray}
  V_{g}(x) &=&
  -\frac{\Delta_0}{3}-
  V_{0}~\vartheta(|x|-h-a)+
  V_{d}~\vartheta(h-|x|)
  \nonumber \\ && +
  V_{b}~\vartheta(|x|-h)~\vartheta(h+a-|x|),
  \label{V-gate-def}
\end{eqnarray}
where $V_b>V_d>0$ and $V_0>0$. This gate divides the CNT
into 5 intervals whose geometry and energy levels are displayed
in Figs~\ref{Fig1a} and \ref{Fig1b}:\\
1) $|x|>h+a$, (left and right leads): Here the Fermi energy
      $\epsilon_F$ is over the bottom of the conduction band.\\
2) $h+a>|x|>h$, the left and right barriers.\\
3) $|x|<h$, (the QD): Here the Fermi energy is below
      the single-electron level, so that the ground state
      of the QD hosts solely the spin 1/2 caged atom while
      its excited states contain one electron and the caged
      atom. Together, they form singlet (S) and triplet (T)
      states. The exchange interaction between the atom
      and the electron is considered in the SM
       (based on Hund rules, as it is quite a standard
      aspect of atomic physics). It is shown that the singlet-triplet
      splitting $V_{\mathrm{exch}}=\epsilon_T-\epsilon_S$ due to the direct
      exchange term is $V_{\mathrm{exch}}=-43.3$~meV, while
      the indirect exchange term is absent. Since
      $V_{\mathrm{exch}}<0$ the exchange interaction is
      {\it ferromagnetic} and the corresponding energies satisfy
      $\epsilon_S>\epsilon_T$, see Fig.~\ref{Fig1b}.

\noindent
\underline{\it Notations for Anderson model:} 
1) $c_{\alpha q \xi \sigma}$ are electron annihilation operators
in the leads $\alpha=L,R$, with momentum $q$, energy
$\epsilon_q=\sqrt{(\hbar v q)^2+\tfrac{1}{9}\Delta_0^2}$,
flavour $\xi$=${\bf K},{\bf K}'$ and spin projection
$\sigma$=$\pm \tfrac{1}{2}$.  2) For a lead of length $L$ the 
electron DOS is
$\rho(\epsilon)=\tfrac{1}{L}\sum_{k}\delta(\epsilon-\epsilon_k)=
\frac{\vartheta(\epsilon-\frac{1}{3}\Delta_0)}{\pi\hbar{v}_{\epsilon}}$
where
$v_\epsilon=\tfrac{v}{\epsilon} \sqrt{\epsilon^2-\tfrac{1}{9}\Delta_0^2}$
is the group velocity.
3) $d_{\xi \sigma}$ is
the annihilation operator for dot electron of flavour $\xi$ and
spin projection $\sigma$.  4) $|M_A \ra$ is a dot atom
state with spin projection $M_A$=$\ua,\da$ and energy $\epsilon_{M_A}$=$0$.
5) The atom (doublet) and atom-electron singlet and three  triplet
states of the dot are collectively denoted as
$|\Lambda\ra$=$|M_A\ra$, $|S\xi\ra$, $|1\xi\ra, |0\xi\ra$, $|\bar{1}\xi\ra$.
The triplet states are,\\
$|1\xi\rangle$=$d_{\xi\ua}^{\dag}|\ua\rangle$,
$|0\xi\rangle$=$\tfrac{1}{\sqrt{2}}
 \{d_{\xi\ua}^{\dag}|\da\rangle$+$
   d_{\xi\da}^{\dag}|\ua\rangle\}$,
$|\bar{1}\xi\rangle$=$
 d_{\xi\da}^{\dag}|\da\rangle,$
and the singlet state is
$|S\xi\rangle$=$\frac{1}{\sqrt{2}}
 \{d_{\xi\ua}^{\dag}|\da\rangle-
   d_{\xi\da}^{\dag}|\ua\rangle\}$.\\
6)   The dot Hubbard operators  are 
$X^{\Lambda,\Lambda'} \equiv |\Lambda\ra\la\Lambda'|$. 
The dot electron operators $d_{\xi \sigma}$  defined
in 3) are expressible in terms of 
$X^{M_A,S \xi}, X^{M_A,1 \xi}, X^{M_A,0 \xi},X^{M_A,\bar{1} \xi}$ as,\\
$d_{\xi \sigma}=X^{\sigma,1\xi}+\frac{1}{\sqrt{2}}
\big\{X^{\bar{\sigma},0\xi}+2\sigma X^{\bar{\sigma},S\xi}\big\}$,
($\bar{\sigma} \equiv - \sigma$).

\noindent
\underline{\it Anderson and Kondo Hamiltonians} The Anderson Hamiltonian of
the CNTL-CNTQDA-CNTR junction is:
\begin{subequations}
\begin{eqnarray}
  &&
  H =
  H_{L}+H_{R}+H_{D}+H_{T},
  \label{H-Anderson-def}
  \\
  &&
  H_{\alpha} =
  \sum_{q \xi \sigma}
  \epsilon_q
  c_{\alpha q \xi \sigma}^{\dag}
  c_{\alpha q \xi \sigma},
  \label{H-LR-def}
  \\
  &&
  H_{D} =
  \sum_{\xi}
  \Big\{
      \epsilon_{T}
      \sum_{m}
      X^{m{\xi},m{\xi}}+
      \epsilon_{S}
      X^{S{\xi},S{\xi}}
  \Big\},
  \label{H-D-def}
  \\
  &&
  H_{T} =
  \sum_{\alpha q \xi \sigma}
  t_{\epsilon_q}~
  \Big\{
      c_{\alpha q \xi \sigma}^{\dag}
      d_{\xi\sigma}+
      d_{\xi\sigma}^{\dag}
      c_{\alpha q \xi \sigma}
  \Big\},
  \label{H-T-def}
\end{eqnarray}
  \label{subeqs-H-Anderson}
\end{subequations}
Here $t_{\epsilon}=t_F\sqrt{\frac{v_{\epsilon}}{v_F}}$,
$v_F=v_{\epsilon_F}$, and $t_F$ is
the tunneling rate for electrons at the Fermi level.
The energies of the triplet and singlet states,
$\epsilon_{T}$ and $\epsilon_{S}$, satisfy the properties,
$\epsilon_{S}>\epsilon_{T}>\epsilon_{F}$,
(see Fig.~\ref{Fig1b}). 
Applying the Schrieffer-Wolf transformation leads to
the Kondo Hamiltonian
\begin{eqnarray}
&&  H_K =
  \sum_{\xi q q'}
  \Big\{
      K_{\epsilon_{q}\epsilon_{q'}}
      n_{\xi,q q'}+
      J_{\epsilon_{q}\epsilon_{q'}}
      \big(
          {\mathbf{S}}
          \cdot
          {\mathbf{s}}_{\xi,q q'}
      \big)
  \Big\},    \label{HK-res} 
  \\
  &&
  n_{\xi, q q'} ~=~
  \sum_{\alpha \alpha' \sigma}
  c_{\alpha q \xi \sigma}^{\dag}
  c_{\alpha' q' \xi \sigma}, \nonumber
  \\
  &&
  {\mathbf{s}}_{\xi, q q'} ~=~
  \frac{1}{2}
  \sum_{\alpha \alpha' \sigma \sigma'}
  c_{\alpha q \xi \sigma}^{\dag}
  {\boldsymbol\tau}_{\sigma\sigma'}
  c_{\alpha' q' \xi \sigma'}, \nonumber
  \\
  &&
  {\mathbf{S}} ~=~
  \frac{1}{2}
  \sum_{\sigma\sigma'}
  {\boldsymbol\tau}_{\sigma\sigma'}
  X^{\sigma\sigma'}. \nonumber
  \end{eqnarray}
The couplings $K_{\epsilon\epsilon'}$ and $J_{\epsilon\epsilon'}$
are
\begin{eqnarray}
  K_{\epsilon\epsilon'} &=&
  \frac{3t_{\epsilon}t_{\epsilon'}}
       {4(\epsilon_T-\epsilon_F)}+
  \frac{t_{\epsilon}t_{\epsilon'}}
       {4(\epsilon_S-\epsilon_F)},
  \nonumber \\
  J_{\epsilon\epsilon'} &=&
  \frac{t_{\epsilon}t_{\epsilon'}}
       {\epsilon_T-\epsilon_F}-
  \frac{t_{\epsilon}t_{\epsilon'}}
       {\epsilon_S-\epsilon_F}.
  \label{exchange-couplings}
\end{eqnarray}
In the special case $\epsilon_S=\epsilon_T$,  $H_K$ contains
just a potential scattering which does not include spin-flipping.
Since $\epsilon_T<\epsilon_S$, an anti-ferromagnetic spin-spin
exchange interaction appears ($J>0$).
Eq. (\ref{HK-res}) describes two-channel Kondo scattering
with dot spin $S=\tfrac{1}{2}$, so that it results in
over-screening of the impurity spin.

\noindent
\underline{ \it Scaling equation and Kondo temperature:}
Employing the poor man's scaling technique to the Kondo Hamiltonian
(\ref{HK-res}), one can see that only the dimensionless
coupling
$j_{\epsilon\epsilon'}=J_{\epsilon\epsilon'}
 \sqrt{\rho(\epsilon)\rho(\epsilon')}=
  \frac{t_F^2 (\epsilon_S-\epsilon_T)
         \vartheta(\epsilon-\frac{\Delta_0}{3})
         \vartheta(\epsilon'-\frac{\Delta_0}{3})}
       {\pi\hbar{v}_{F}(\epsilon_S-\epsilon_F)(\epsilon_T-\epsilon_F)}$
renormalizes.
Note that for $\epsilon,\epsilon'>\frac{1}{3}\Delta_0$,
${j}\equiv{j}_{\epsilon\epsilon'}$ does not depend on $\epsilon$
and $\epsilon'$.

\begin{figure}[htb]
\centering
 \includegraphics[width=50 mm,angle=0]
    {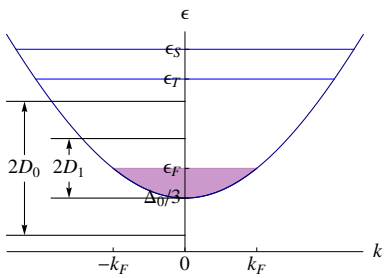}
 \caption{\footnotesize
  (Color online)
  Two different intervals of the effective bandwidth $D$,
  $D_0<D<D_1$ and $D<D_1$, where different
  RG regimes are expected.}
  \vspace{-0.15in}
 \label{Fig-regimes}
\end{figure}
The RHS of the scaling equation (see below) consists of
the following terms,
$
  j_{\epsilon\epsilon+D}
  j_{\epsilon+D\epsilon'}$,
 and
$ j_{\epsilon\epsilon'-D}
  j_{\epsilon-D\epsilon'},
$
where $\epsilon$ and $\epsilon'$ are close to $\epsilon_F$.
The first term does not depend on energies and
equals to $j^2$, whereas the second term depends on
the sign of $\epsilon_F-D-\frac{1}{3}\Delta_0$.
For $\epsilon_F-D>\frac{1}{3}\Delta_0$, 
$j_{\epsilon\epsilon-D}j_{\epsilon-D\epsilon'}= j^2$, otherwise
$j_{\epsilon\epsilon-D}j_{\epsilon-D\epsilon'}=0$. As a result,
for $D_0>\epsilon_F-\frac{1}{3}\Delta_0$, there are two different
regimes of the poor man's scaling procedure:
$D_0<D<D_1=\epsilon_F-\frac{1}{3}\Delta_0$ and $D<D_1$,
as shown in Fig. \ref{Fig-regimes}. Within the first of them,
$D_0<D<D_1$, the energy level $\epsilon_F+D$ lies
within the conduction band, whereas the energy
$\epsilon_F-D$ is below the bottom of the conduction
band. Within the second interval, $D<D_1$, both
the energy levels, $\epsilon_F\pm{D}$, lie within
the conduction band. This property effects on
the RG procedure.We will consider the RG procedure within both of
the intervals in turn.

\noindent
{\textbf{First interval}}, $D_0<D<D_1$:
The second and third order diagrams containing hole
lines (i.e., lines describing virtual electrons with energy
$\epsilon-D$) vanish. Then scaling equations for $k$ and $j$ are,
\begin{subequations}
\begin{eqnarray}
  \frac{\partial{k}}
       {\partial\ln{D}}
  &=&
  -k^2-
  \frac{3j^2}{16},
  \label{scale-eq-k-D>D1}
  \\
  \frac{\partial{j}}
       {\partial\ln{D}}
  &=&
  -2kj-
  \frac{j^2}{2},
  \label{scale-eq-j-D>D1}
\end{eqnarray}
  \label{subeqs-scale-eqs-D>D1}
\end{subequations}
where $D<D_0$ and the initial conditions are
$k(D_0)=k_0$ and $j(D_0)=j_0$.

The solution of the set of equations (\ref{subeqs-scale-eqs-D>D1}) is,
\begin{subequations}
\begin{eqnarray}
  k(D) &=&
  \frac{1}
       {\displaystyle
        4\ln
        \bigg(
            \frac{D}{T_g}
        \bigg)}+
  \frac{3}
       {\displaystyle
        4\ln
        \bigg(
            \frac{D}{T_f}
        \bigg)},
  \label{k(D)-D>D1-res}
  \\
  j(D) &=&
  \frac{1}
       {\displaystyle
        \ln
        \bigg(
            \frac{D}{T_g}
        \bigg)}-
  \frac{1}
       {\displaystyle
        \ln
        \bigg(
            \frac{D}{T_f}
        \bigg)},
  \label{j(D)-D>D1-res}
\end{eqnarray}
  \label{subeqs-scale-solution-D>D1}
\end{subequations}
where
\begin{eqnarray*}
  T_g ~=~
  D_0\exp\Big(-\frac{1}{g_0}\Big),
  \ \ \
  T_f ~=~
  D_0\exp\Big(-\frac{1}{f_0}\Big),
\end{eqnarray*}
$g_0=k_0+\frac{3}{4}j_0$ and $f_0=k_0-\frac{1}{4}j_0$.
Since $g_0>f_0$, $T_g>T_f$.

Provided that $D_1>T_g$, the renormalization procedure
(\ref{subeqs-scale-solution-D>D1}) stops when $D$ approaches
$D_1$ and $k$ and $j$ approach $k_1=k(D_1)$ and $j_1=j(D_1)$.
From this point, the second RG regime starts.

\begin{figure}[htb]
\centering
 \includegraphics[width=55 mm,angle=0]
    {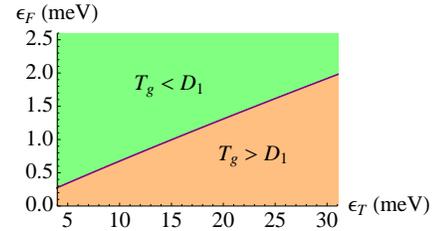}
 \caption{\footnotesize
  (Color online)
  Intervals of $\epsilon_T$ and $\epsilon_F$ where
  $T_{K_1}<D_1$ (green area) or
  $T_{K_1}>D_1$ (orange area)
  separated by the purple curve
  where $T_{K_1}=D_1$. Here $\epsilon_S-\epsilon_T=43.3$~meV.}
 \label{Fig-PhD}
\end{figure}

Intervals of $\epsilon_F$ and $\epsilon_T$ where
$T_g<D_1$ or $T_g>D_1$ are shown in
Fig. \ref{Fig-PhD}. It is seen that there are some
values $\epsilon_c(\epsilon_T)$ (purple curve)
such that for $\epsilon_F>\epsilon_c(\epsilon_T)$,
$T_g<D_1$ and the crossover from
the single-channel to the two-channel Kondo
regimes occurs in the weak coupling regime.
When $\epsilon_F<\epsilon_c(\epsilon_T)$,
$T_g>D_1$ and the crossover from
the single-channel to the two-channel Kondo
regimes occurs in the strong coupling regime.

\noindent
{\textbf{Second interval}}, $D<D_1$:
The third order scaling equation for $j$ is,
\begin{eqnarray}
  \frac{\partial{j}(D)}
       {\partial\ln{D}}
  &=&
  -j^{2}(D)+
  2j^{3}(D),
  \label{scal-eq-third-order-D<D1}
\end{eqnarray}
and the initial condition is $j(D_1)=j_1$.
Within this interval, the coupling $k$ does
not renormalizes. The solution of the scaling equation
(\ref{scal-eq-third-order-D<D1}) is,
\begin{eqnarray}
  \frac{1}{j_1}-
  \frac{1}{j}+
  2\ln
  \bigg[
       \frac{j(1-2j_1)}
            {j_1(1-2j)}
  \bigg]
  &=&
  \ln\frac{D_1}{D}.
  \label{solution-scal-eq-D<D1}
\end{eqnarray}
When $D$ decreases, $j(D)$ renormalizes towards
$j^{*}=\frac{1}{2}$, the fixed point value of $j$, and when $D$
goes to $0$, $j(D)$ goes to $j^{*}$. When $|j-j^{*}|\ll{j}^{*}$,
the asymptotic expression for $j$ is,
\begin{eqnarray*}
  \frac{j-j^{*}}{j^{*}} &=&
  \frac{j^{*}-j_1}{j_1}~
  \bigg(
       \frac{DT^{*}}{D_1T_K}
  \bigg)^{j^{*}},
\end{eqnarray*}
where the scaling invariants $T_K$ and $T^{*}$ are
\begin{eqnarray}
  T_K =
  D_1
  \exp
  \bigg(
       -\frac{1}{j_1}
  \bigg),
  \ \ \
  T^{*} =
  D_1
  \exp
  \bigg(
       -\frac{1}{j^{*}}
  \bigg).
  \label{TK}
\end{eqnarray}

\noindent
{\underline{\it Estimate of the couplings and Kondo temperature $T_K$}:
The dimensionless couplings are,
\begin{eqnarray*}
  j_0 &=&
  \frac{t_F^2 (\epsilon_S-\epsilon_T)}
       {\pi\hbar{v}_{F}
        (\epsilon_S-\epsilon_F)
        (\epsilon_T-\epsilon_F)},
  \\
  k_0 &=&
  \frac{t_F^2 (3\epsilon_S+\epsilon_T-4\epsilon_F)}
       {4\pi\hbar{v}_{F}
        (\epsilon_S-\epsilon_F)
        (\epsilon_T-\epsilon_F)}.
\end{eqnarray*}
Here $\epsilon_S-\epsilon_T=43.3$~meV (see the SM). The tunneling
rate can be tuned by fitting the tunnel barrier height and width.

\begin{figure}[htb]
\centering
\centering\subfigure[]{\includegraphics[width=0.23\textwidth]
{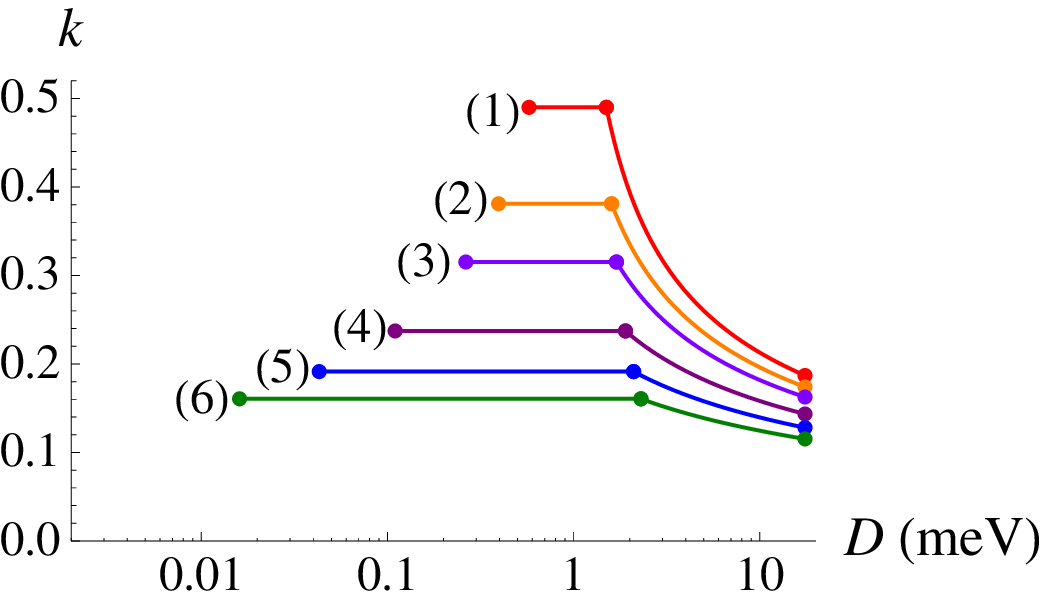}
\label{Fig-k(D)}}
\centering\subfigure[]{\includegraphics[width=0.23\textwidth]
{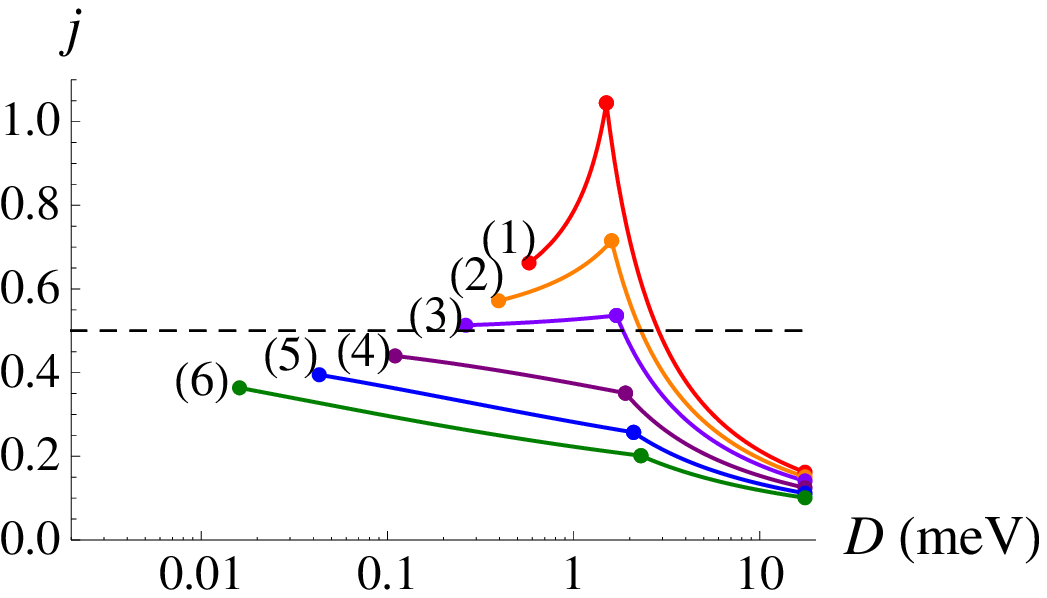}
\label{Fig-j(D)}}
 \caption{\footnotesize
  (Color online)
  (a)  $k$ and (b) $j$ as functions of $D$ for
  $\epsilon_T=19$~meV
  and different values of $\epsilon_F$.
  Here $\epsilon_S-\epsilon_T=43.3$~meV,
  $t_F\sqrt{k_F}=0.35(\epsilon_T-\epsilon_F)$ and
 curves (1) -- (6) correspond to
  $\epsilon_F$=1.5, 1.6, 1.7, 1.9, 2.1 and 2.3~meV,
  respectively.}
 \label{Fig-k(D)-j(D)}
\end{figure}
The dependence of the effective couplings $k$ and $j$ on
the effective bandwidth $D$ and the Fermi energy $\epsilon_F$
is shown in Fig.\ref{Fig-k(D)-j(D)} for the energy of the triplet state
$\epsilon_T=19$~meV. $k(D)$ as a function of $D$ is
shown in Fig.\ref{Fig-k(D)} and renormalization of $j(D)$
is shown in Fig.\ref{Fig-j(D)} for different values of $\epsilon_F$.
It should be noted the behavior of the curves (1), (2) and (3)
[$\epsilon_F\leq1.7$~meV]:
Within the interval $D_0>D>D_1$, the effective coupling
$j(D)$ increases to the value over $j^{*}$, and then within
the interval $D<D_1$, $j(D)$ decreases approaching $j^{*}$.
This behavior is unexpected, since in the standard
two-channel Kondo model, the exchange coupling changes
monotonically with $D$ approaching $j^{*}$ for $D\to0$.
The nonmonotonic behavior is caused by the crossover
from the single-channel GR regime for $D>D_1$ to
the two-chanel RG regime for $D<D_1$.

\begin{figure}[htb]
\centering
\centering\subfigure[]{\includegraphics[width=0.235\textwidth]
{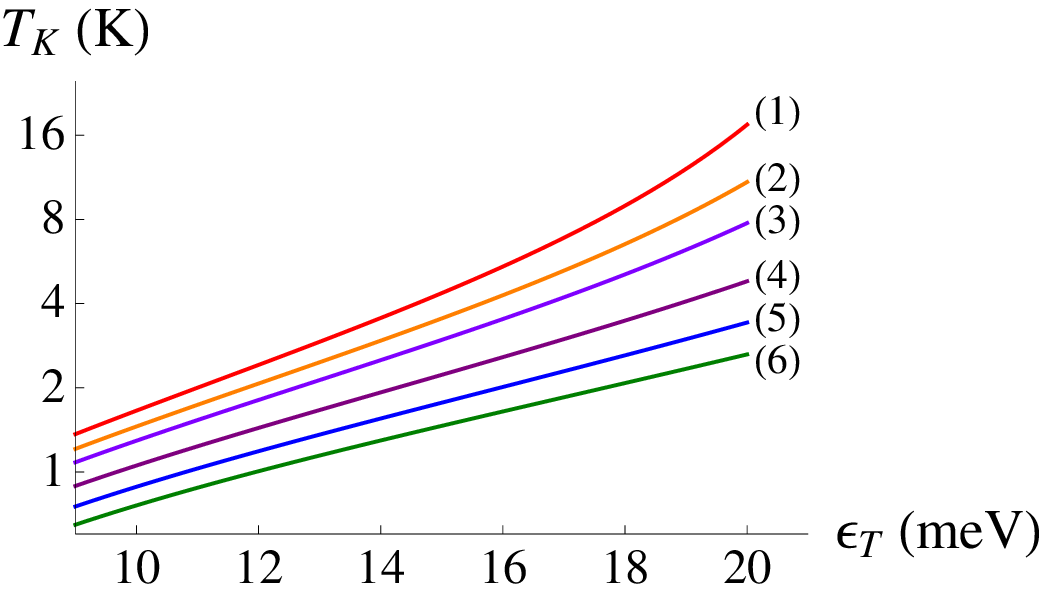}}
%
\centering\subfigure[]{\includegraphics[width=0.235\textwidth]
{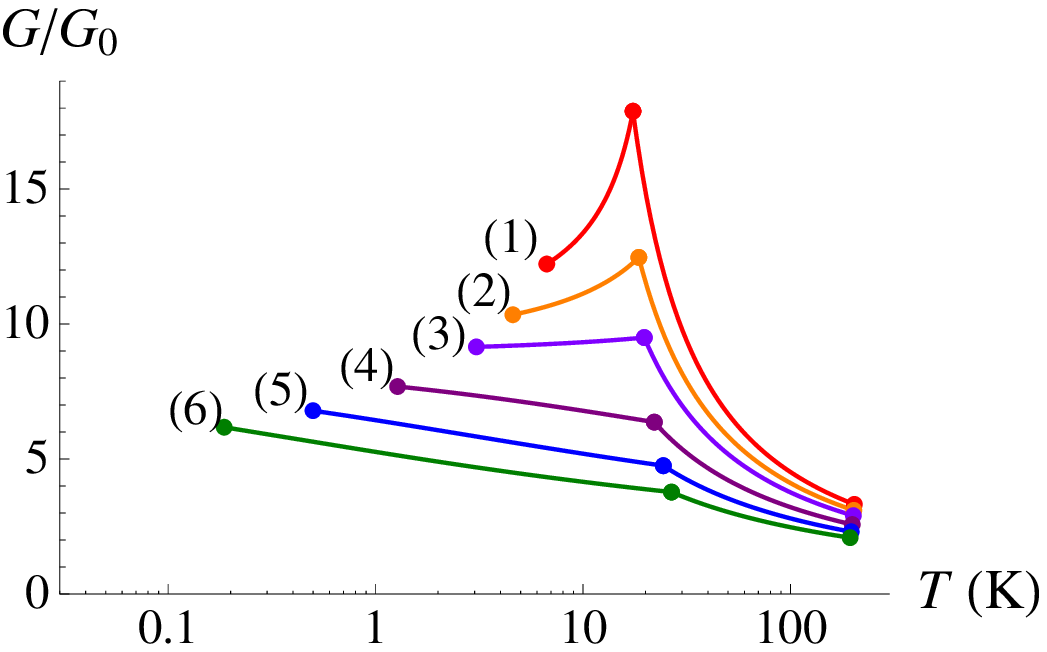}}
 \caption{\footnotesize
  (Color online)
  (a)  $T_K$, Eq. (\ref{TK}), as a function of $\epsilon_T$
  and different values of $\epsilon_F$.
  (b) The conductance $G(T)$, Eq.~(\ref{G-poor-mans}) as function of $T$ for
  $\epsilon_T=19$~meV and different values of
  $\epsilon_F$. 
  For both panels, $t_F\sqrt{k_F}=0.35(\epsilon_T-\epsilon_F)$,  
  and curves (1) -- (6) correspond to
  $\epsilon_F$=1.5, 1.6, 1.7, 1.9, 2.1 and 2.3~meV,
  respectively.
  In panel (b), the dots from right to left correspond to
  $D_0, D_1$ and
  $T_K$, separating the RG regimes
  from one another.}
  \vspace{-0.15in}
 \label{Fig-TKG}
\end{figure}
The Kondo temperature $T_K$ (\ref{TK}) is shown in Fig.~
\ref{Fig-TKG}a as a function of $\epsilon_T$ for
$t_F\sqrt{k_F}=0.35(\epsilon_T-\epsilon_F)$ and
different values of $\epsilon_F$. It is seen that $T_K$ changes in
between $0.5$~K and $5$~K for reasonable parameter values.

\noindent
\underline{\it Conductance}
The non-linear tunneling conductance $G(T)$
of the CNTL-CNTQDA-CNTR junction in the weak coupling
Kondo regime will now be calculated, employing perturbation RG formalism
within the Keldysh non-equilibrium Green's function approach.
The tunneling current from the left
to the right lead is
\begin{eqnarray}
  I &=&
  \frac{i e}{\hbar}
  \sum_{\xi k k' \sigma \sigma'}
  \Big\{
      K_{\epsilon_{k}\epsilon_{k'}}
      \delta_{\sigma\sigma'}+
      \frac{J_{\epsilon_{k}\epsilon_{k'}}}{2}~
      \big(
          \boldsymbol\tau_{\sigma\sigma'}
          \cdot
          {\mathbf{S}}
      \big)
  \Big\}
  \nonumber \\ && \times
  \Big(
      c_{L k \xi \sigma}^{\dag}
      c_{R k' \xi \sigma'}-
      c_{R k \xi \sigma}^{\dag}
      c_{L k' \xi \sigma'}
  \Big).
  \label{current-def}
\end{eqnarray}
Applying perturbation theory and the condition of invariance of
the conductance with respect to the ``rescaling'' transformations,
we get the following expression for the conductance,
\begin{eqnarray}
  G =
  \frac{\pi^2 G_0}{2}~
  \Big\{
      k^2(T)+
      3j^2(T)
  \Big\},
  \ \ \
  G_0=\frac{e^2}{\pi\hbar},
  \label{G-poor-mans}
\end{eqnarray}
where $j(T)$ is given by Eqs. (\ref{j(D)-D>D1-res}) and
(\ref{solution-scal-eq-D<D1}) for $T>D_1$
and $T<D_1$, respectively; $k(T)$ is given by
Eq. (\ref{k(D)-D>D1-res}) for $T>D_1$ and
$k(T)=k(D_1)$ for $T<D_1$ [see also Fig. \ref{Fig-k(D)-j(D)}].

The conductance (\ref{G-poor-mans}) as function of $T$
is shown in Fig. \ref{Fig-TKG}b for
$t_F\sqrt{k_F}=0.35(\epsilon_T-\epsilon_F)$,
$\epsilon_T=19$~meV and different values of $\epsilon_F$.
Note the non-monotonic behavior of the conductance for
$\epsilon_F\leq1.7$~meV [curves (1) -- (3)]. This
exotic behavior is caused by the non-monotonicity of
$j(T)$ [see Fig. \ref{Fig-j(D)}]. 
In the standard 2CKE, $G(T)$ is monotonic, 
depending on the bare value $j_0$ of $j$. If $j_0<j^{*}=\frac{1}{2}$, ($j_0>j^{*}$), 
the conductance increases (decreases) monotonically with reducing
$T$. Non-monotonicity of $G(T)$ exposed here
is the result of the crossover between different
RG scaling regimes.\\
\underline{\it Summary:} In addition to suggesting a scheme for 
exposing 2CKE in an electronic transport system, the present scheme reveals a novel facet of 
the RG scaling framework, namely, the existence of two scaling regimes in which the 
running coupling constant behaves differently. This enables the physics of the 2CKE to be 
visible also in the weak coupling regime due to the non-monotonic behaviour of the conductance as function of temperature. 
Our analysis requires a versatile  use of several physical disciplines.
The properties of CNT are employed for the generation of decoupled electron channels, while 
the need to implant an atom on the axis of the CNT  requires mastering of material science techniques. 
The elucidation of the van der Waals potential and the CNT and the calculation of the exchange constants 
are based on fundamental aspects of atomic physics, and the derivation of the Anderson and Kondo Hamiltonians 
touch upon the corner sotnes of strongly correlated electrons. \\
\underline{\it Acknowledgement:} The research of Y.A is partially supported by grant 400/12 of the Israel Science Foundation (ISF).

\onecolumngrid

\section{Supplementary Material}
In this Supplementary Material section we first expand upon
the atomic-physics aspects of the  CNTL-CNTQDA-CNTR
junction and second, derive the scaling equations for
the exchange coupling constants. 
In subsection I the van der Waals interaction
between the gold atom and the CNT is derived based
on the van der Waals interaction between the gold atom
and a single carbon atom and on the geometry of the CNT.  
In subsection II the gold atom states in the CNT potential
(derived in subsection I) are analyzed.
Direct exchange interaction between the CNT itinerant
electrons and the caged gold atom are discussed in
subsection III providing us with the exchange constants
that are required to arrive at the Kondo model of the main
text. In subsection IV it is shown that the indirect exchange
contribution is virtually negligible.
Finally, in subsection V the scaling equations for
the dimensionless couplings are derived. The reason for doing
so is that these equations are not the standard ones encountered
in the Kondo physics because, as indicated in the main text,
there are two different scaling intervals, $D>D_1$ and $D<D_1$,
corresponding to two different sets of scaling equations, and
each regions requires a separate treatment.

\subsection{I. Van der Waals Interaction between gold and
           carbon atoms forming CNT}

When an atom of gold is implanted on the central axis of
a CNT,  there is van der Waals interaction between the  gold
and {\it all} the carbon atoms forming the CNT that need to be
calculated. In order to calculate it we first write down 
the van der Waals interaction between the gold and
a {\it single} carbon atom a distance $Y$ apart,  that reads,  
\begin{eqnarray}
  V_{\mathrm{w}}(y) &=&
  V_0~
  \bigg\{
       \frac{1}{2}~
       \frac{R_{\mathrm{w}}^{12}}{Y^{12}}-
       \frac{R_{\mathrm{w}}^{6}}{Y^{6}}
  \bigg\},
  \label{van-der-Waals-atom-atom}
\end{eqnarray}
where $R_{\mathrm{w}}=3.36$~{\AA} is the equilibrium position and
\begin{eqnarray*}
  V_0 &=&
  \frac{\veps_{\mathrm{Au}}\veps_{\mathrm{C}}}
       {\veps_{\mathrm{Au}}+\veps_{\mathrm{C}}}~
  \frac{\alpha_{\mathrm{Au}}\alpha_{\mathrm{C}}}
       {r_{\mathrm{w}}^{6}}
  ~=~
  32.6~{\mathrm{meV}},
\end{eqnarray*}
where $\veps_{\mathrm{Au}}=9.2255$~eV or
$\veps_{\mathrm{C}}=11.2603$~eV is the ionization energy for gold
or carbon, $\alpha_{\mathrm{Au}}=36.1a_B^3=5.34$~{\AA}$^{3}$ or
$\alpha_{\mathrm{C}}=11.7a_B^3=1.73$~{\AA}$^{3}$ is polarizability
of the atom of gold or carbon ($a_B=0.529$~{\AA} is the Bohr
radius).
Then the van der Vaals interaction of the caged atom of gold with the
CNT (see Fig.~\ref{Fig-CNT-Au} for the geometry) is obtained just by summation,
\begin{eqnarray}
  V({\mathbf{R}}) &=&
  \sum_{{\mathbf{R}}_{A}}
  V_{\mathrm{w}}\big(\big|{\mathbf{R}}-{\mathbf{R}}_{A}\big|\big)+
  \sum_{{\mathbf{R}}_{B}}
  V_{\mathrm{w}}\big(\big|{\mathbf{R}}-{\mathbf{R}}_{B}\big|\big),
  \label{van-der-Waals-atom-CNT}
\end{eqnarray}
where ${\mathbf{R}}_{A}$ or ${\mathbf{R}}_{B}$ are positions of
atoms of sub-lattice $A$ or $B$.
Numerical calculations for $V({\mathbf{R}})$ result its profile as displayed in Fig. 2 of
the main text for a zigzag CNT with the chiral vector
${\mathbf{c}}_{8,0}$ and CNT radius $r_0=3.13$~{\AA}. Essentially, the potential
depends on two variables (cylindrical coordinates), $X$ (along the CNT axis) and $R$ (radial
variable) and almost does not depend on the azimuthal angle
$\phi$. Hence we may write $V({\mathbf{R}})=V(R,X)$, which is, by construction, periodic in $X$. 
More concretely, for fixed $X$ it increases quadratically with $R$, and for $R=0$, $V(0,X)$ as a function of
$X$ has local minima $X_n^{\mathrm{min}}$ and saddle points $X_n^{\mathrm{sad}}$ given by,
\begin{eqnarray}
  X_n^{\mathrm{min}} ~=~
  2nX_1,
  \ \ \ \ \
  X_n^{\mathrm{sad}} ~=~
  (2n+1)X_1,
  \ \ \ \ \
  X_1 ~=~
  \frac{3 a_0}{4\sqrt{3}}
  ~=~
  1.07~{\text{\AA}},
  \label{x-min-max}
\end{eqnarray}
where $n$ is integer.
Numerical estimates show that neighboring minima are separated
by tunnel barriers of height
\begin{eqnarray}
  W_b &=&V(0,X_n^{\mathrm{sad}})=
  32.8~{\mathrm{meV}}.
  \label{V-barrier-append}
\end{eqnarray}
 Expansion of $V(R,X)$ around the minima
$\mathbf{R}_n^{\mathrm{min}}=(X_n^{\mathrm{min}},0)$ up to quadratic powers yield, 
\begin{eqnarray}
  V(R,X) &\approx&
  \tfrac{1}{2}K_x
  \big(X-X_n^{\mathrm{min}}\big)^2+
  \tfrac{1}{2}K_r
  R^2,
  \label{van-der-Waals-harmonic}
\end{eqnarray}
where $|X-X_n^{\mathrm{min}}|\ll{X_1}$ and ${R}\ll{r}_{0}$,
\begin{eqnarray}
  K_x ~=~
  0.072~
  \frac{\mathrm{eV}}{{\mathrm{\AA}}^2},
  \ \ \ \ \
  K_r ~=~
  1.117~
  \frac{\mathrm{eV}}{{\mathrm{\AA}}^2}.
  \label{Kx-Kr-data}
\end{eqnarray}
In what follows, we will assume that the gold atom performs small oscillations around 
the point ${\mathbf{R}}=(0,0)$.

\begin{figure}[htb]
\centering
 \includegraphics[width=50 mm,angle=0]
    {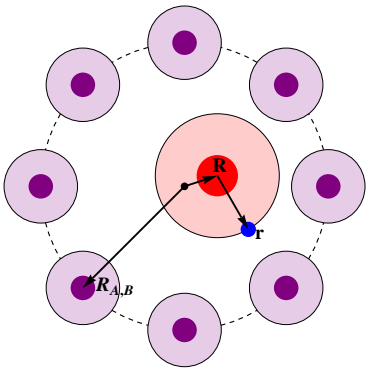}
 \caption{\footnotesize
  (Color online)
  Atom of gold (red circle) and carbon atoms of the CNT
  (purple circles). Position of electron of the gold atom
  is ${\bf{r}}$, position of the gold nuclei is ${\bf{R}}$,
  positions of the carbon atoms are ${\mathbf{R}}_{A}$
  and ${\mathbf{R}}_{B}$.}
 \label{Fig-CNT-Au}
\end{figure}

\subsection{II. Atomic Quantum States in the CNT Potential}

Consider a neutral gold atom of mass $M$  as
a positively charged rigid  ion (with filled shell) and one electron
on the outer 6s orbital. The positions of the ion and the outer
electron are respectively specified by vectors ${\bf{R}}$ and
${\mathbf{r}}$ [see Fig. \ref{Fig-CNT-Au}]. In the adiabatic
approximation (which is natural in atomic physics),
the wave function of the atom is a product of
the corresponding wave functions
$\Psi^{\textrm{Au}}({\bf{R}})$ and $\psi_{6}({\bf{r}})$
describing the stationary states of the ion and
the outer electron. In order to find the wave functions
and energies of the gold atoms in the anisotropic
potential well (\ref{van-der-Waals-harmonic}), we
need to solve the following Schr\"odinger equation for
$\Psi^{\mathrm{Au}}({\bf{R}})$,
\begin{eqnarray}
 -\frac{\hbar^2}{2M} \Delta \Psi^{\mathrm{Au}}({\bf{R}})~
       +V({\bf{R}})
  \Psi^{\mathrm{Au}}({\bf{R}})
  =
  \veps
  \Psi^{\mathrm{Au}}({\bf{R}}).
  \label{eq-Schrodinger-def}
\end{eqnarray}
Within the Harmonic approximation (\ref{van-der-Waals-harmonic}), the solutions of Eq. (\ref{eq-Schrodinger-def}) are,
\begin{subequations}
\begin{eqnarray}
  \Psi_{n m \ell}^{\rm{Au}}({\bf{R}}) &=&
  \Phi_{\nu m}(R)~
  F_{\ell}(X)~
  e^{im\phi},
  \label{Psi-3D-res}
\end{eqnarray}
where $X$, $R$ and $\phi$ are cylindrical coordinates. Denoting
$\rho\equiv R/a_r$, the radial wave function
$\Phi_{nm}(R)$ is,
\begin{eqnarray}
  \Phi_{nm}(R) =
\frac{1}{a_{r}\sqrt{\pi}}~
  \sqrt{\frac{n!}{(n+|m|)!}}~
  \rho^{|m|}
  L_{\nu}^{(|m|)}\big(\rho^2\big)
  e^{\mbox{-}\frac{\rho^2}{2}},
  \label{wave-fun-radial-harm}
\end{eqnarray}
where $L_{\nu}^{(|m|)}$ is the generalized Laguerre polynomial,
$n=0,1,2,\ldots$ and $m=0,\pm1,\pm2,\ldots$, and
\begin{eqnarray}
  a_{r}=
  \sqrt{\frac{\hbar\omega_{r}}{K_{r}}},
  \ \ \ \ \
  \omega_{r} =
  \sqrt{\frac{K_{r}}{M_{\rm{Au}}}}.
  \label{a-s-omega-s-def}
\end{eqnarray}
Denoting $\zeta \equiv X/a_x$ the motion along $X$ is
described by
\begin{eqnarray}
  F_{\ell}(X) &=&
 \frac{1}{\big(\pi a_{x}^2\big)^{\frac{1}{4}}}
  \frac{1}{\sqrt{2^{\ell}~\ell!}}~
  H_{\ell}(\zeta)
  e^{-\zeta^2/2},
  \label{wave-fun-Z-harm}
\end{eqnarray}
where $H_{\ell}$ is the Hermite polynomial, $\ell$ is the harmonic
quantum number, $\ell=0,1,2,\ldots$.
\begin{eqnarray}
  &&
  a_{x}=
  \sqrt{\frac{\hbar\omega_{x}}{K_{x}}},
  \ \ \ \ \
  \omega_{x} =
  \sqrt{\frac{K_{x}}{M_{\rm{Au}}}}.
  \label{a-perp-omega-perp-def}
\end{eqnarray}
  \label{subeqs-wave-fun-slow}
\end{subequations}
The corresponding energy levels depend on two quantum number,
$n=2\nu+|m|$ and $\ell$,
\begin{eqnarray}
  \veps_{n\ell} =
  \hbar\omega_{r}
  \Big(
      n+
      1
  \Big)+
  \hbar\omega_{x}
  \Big(
      \ell+
      \frac{1}{2}
  \Big).
  \label{energy-levels}
\end{eqnarray}
When $\omega_{r}$ and $\omega_{x}$ are incommensurate,
the degeneracy of the level $(n,\ell)$ is $(n+1)(2s+1)$.
For the values of $K_x$ and $K_r$ given by
Eq. (\ref{Kx-Kr-data}), $\hbar\omega_x=1.24$~meV,
$\hbar\omega_r=4.87$~meV, $a_x=0.13$~{\AA}
and $a_r=0.07$~{\AA}. Then the energy of the ground state
is $\veps_{00}=5.49$~meV. The quantum state with $m=5$
has the excitation energy $5\hbar\omega_r=24.35$~meV
which is of order of the ultraviolet cut off.

\subsection{III. Direct Exchange Interaction between CNT QD
            electron and the Caged Atom}

In order to calculate the exchange interaction between the gold
atom and the itinerant electrons in the CNT we neglect the deviation
of the gold atom from the equilibrium position ${\bf R}=(0,0)$. 
Such a small deviation results in a small variation of
the calculated exchange coupling as calculated below.
The exchange interaction is determined mainly by the Coulomb 
repulsion between the CNT itinerant electron and the outer $6s$
electron of the gold atom. These two electrons can for singlet and
triplet states and the corresponding energies are determined by
Hund rules. When the implanted gold atom is placed on the CNT
axis, the singlet-triplet energy splitting
$V_{\mathrm{exch}}=\veps_T-\veps_S$ due to
the direct exchange interaction between the caged atom
and the CNT wall is,
\begin{eqnarray}
  V_{\mathrm{exch}} &=&
  -2\int{d}^{3}{\mathbf{r}}{d}^{3}{\mathbf{r}}'~
  \psi_{\mathrm{CNT}}^{*}({\mathbf{r}})
  \psi_6^{*}({\mathbf{r}}')~
  \frac{e^2}{\big|{\mathbf{r}}-{\mathbf{r}}'\big|}~
  \psi_{\mathrm{CNT}}({\mathbf{r}}')
  \psi_6({\mathbf{r}}).
  \label{V-exch-append}
\end{eqnarray}
Here $\psi_6({\mathbf{r}})$ is the wave function of the 6s-electron
in the atom of gold and $\psi_{\mathrm{CNT}}({\mathbf{r}})$ is
the wave function of electron in the CNT QD. In order to calculate
the latter we need to recall the formalism for calculating electron
wave function in CNT.

A zigzag CNT is specified by two basic vectors  ${\bf{a}}_1$ and
${\bf{a}}_2$, and a chiral vector ${\mathbf{c}}_{N}$,
\begin{eqnarray}
  {\bf{c}}_{N}=N{\bf{a}}_1,
  \label{chiral-vector-def}
\end{eqnarray}
where $N$ is integer and $|{\bf{a}}_1|=|{\bf{a}}_2|=a_0=2.46$~{\AA}.
A CNT is obtained by rolling a 2D graphene sheet such
that the atom at the origin coincides with the atom at
${\bf{c}}_{N}$. Then $|{\bf{c}}_{N}|=2\pi{r_0}$ is the length of
the CNT circumference and $r_0$ is the CNT radius [see
Figure \ref{Fig-CNT-chiral}].

The coordinates of atoms of the sub-lattice $A$ or $B$ are,
\begin{eqnarray*}
  {\mathbf{R}}_{An_1n_2} &=&
  n_1{\mathbf{a}}_{1}+
  n_2{\mathbf{a}}_{2}+
  \frac{{\mathbf{d}}_{1}}{2} ~=~
  \big(X_{An_2},~Y_{n_1+\frac{n_2}{2}}\big),
  \nonumber \\
  {\mathbf{R}}_{Bn_1n_2} &=&
  n_1{\mathbf{a}}_{1}+
  n_2{\mathbf{a}}_{2}-
  \frac{{\mathbf{d}}_{1}}{2} ~=~
  \big(X_{Bn_2},~Y_{n_1+\frac{n_2}{2}}\big),
\end{eqnarray*}
where
\begin{eqnarray}
  &&
  X_{An_2} ~=~
  \frac{a_0}{2\sqrt{3}}+
  \frac{3n_2a_0}{2\sqrt{3}},
  \ \ \ \ \
  X_{Bn_2} ~=~
  -\frac{a_0}{2\sqrt{3}}+
  \frac{3n_2a_0}{\sqrt{3}},
  \ \ \ \ \
  Y_{n} ~=~
  na_0.
  \label{X-X-Y-def}
\end{eqnarray}
Here $x$-axis is along the CNT axis and the $y$-axis is in the
circumference direction. Then $n_2$ changes from 1 to $N$
and we have the periodicity condition $Y_{n+N}=Y_{n}$.
For the CNT to be semiconductor, $N$ is not an integer multiplier
of 3. We consider here the CNT with  $N=8$ and $r_0=3.13$~nm.

\begin{figure}[htb]
\centering
 \begin{tabular}{|l|l|}
 \hline
 ~ (a) & ~ (b)
 \\
 ~
 \includegraphics[height=45 mm,angle=0]
    {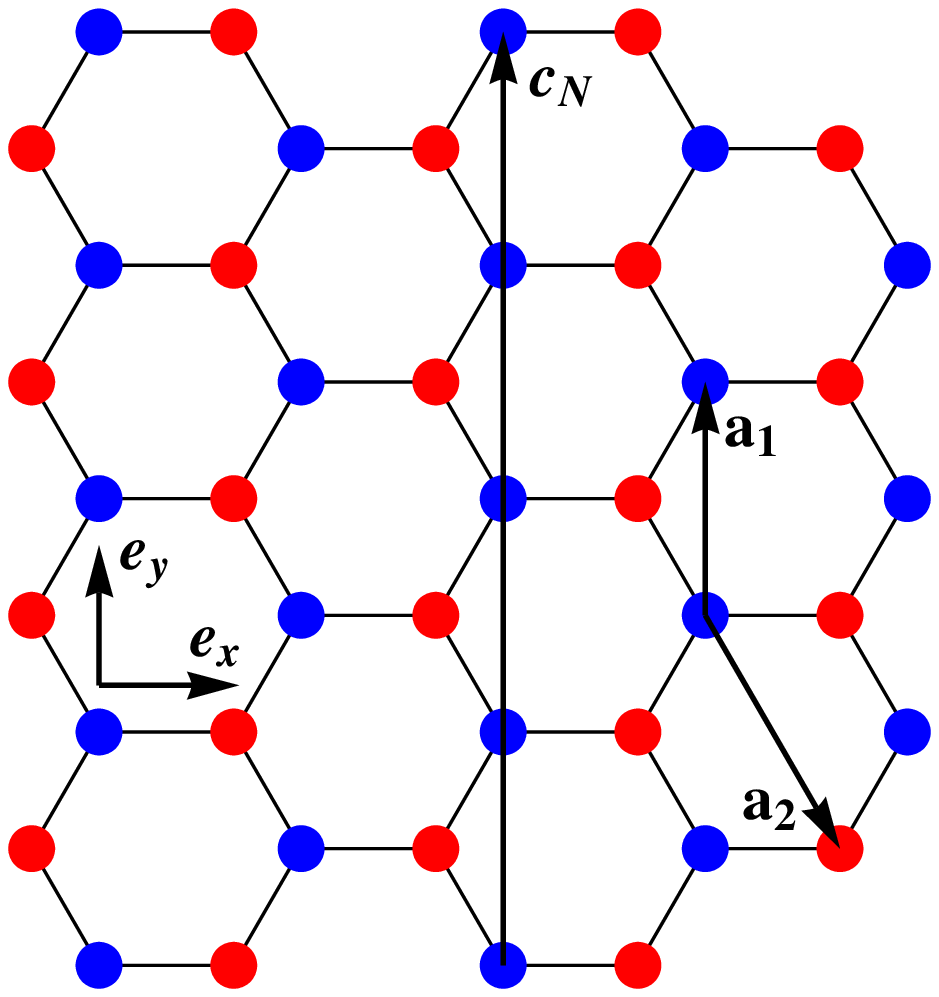}
 ~
 &
 ~
 \includegraphics[height=45 mm,angle=0]
    {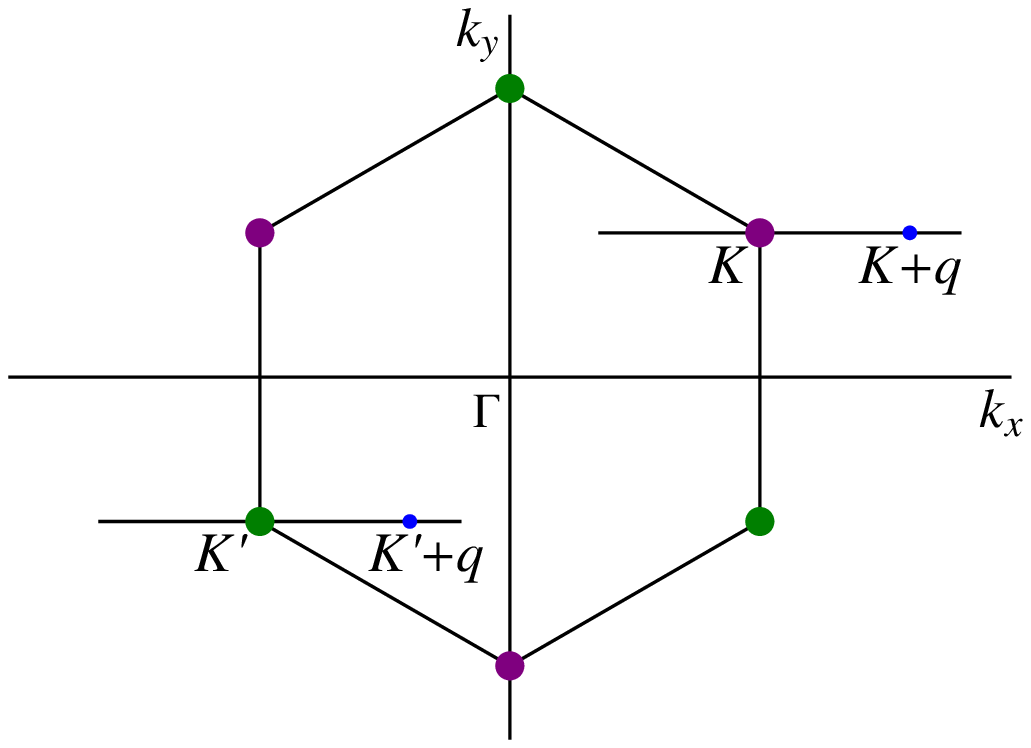}
 \\
 \hline
 \end{tabular}
 \caption{\footnotesize
  (Color online)
  {\bf{Panel (a)}}: A monoatomic layer of graphene. The red and
  blue dots denote carbon atoms of the sub-lattice A and B. The
  primitive vectors of graphene are ${\bf{a}}_1$ and ${\bf{a}}_2$.
  The nanotube is obtained by choosing the chiral vector
  ${\bf{c}}_{N}=N{\mathbf{a}}_1$.
  The unit vectors ${\bf{e}}_x$ and ${\bf{e}}_y$ are fixed in
  the CNT in such a way that ${\bf{e}}_x$ is along the CNT
  axis, and ${\bf{e}}_y$ is along the circumferential direction
  ${\bf{c}}_{N}$.
  {\bf{Panel (b)}}: The first Brillouin zone of graphene.
  $k_x$ is the component of the 2D wave vector ${\mathbf{k}}$ along
  the CNT axis and $k_y$ is the component of ${\bf{k}}$ in
  the circumferential direction. The green and purple
  dots denote the corners ${\bf{K}}$ and ${\bf{K}}'$
  of the first BZ. The lines $q$ denote the component of
  the wave vector ${\mathbf{k}}$ along the CNT axis measured
  with respect to the corners ${\bf{K}}$ and ${\bf{K}}'$.}
 \label{Fig-CNT-chiral}
\end{figure}

The wave function of the lowest-energy quantum state of the CNT QD
of the length $2h$ (i.e., for $|{X}_{A,Bn_2}|\leq{h}$) is
\begin{eqnarray}
  \psi_{\mathrm{CNT}}({\mathbf{r}})
  &=&
  \frac{1}{\sqrt{N}}
  \sum_{n_1n_2}
  \Phi\big({\mathbf{r}}-{\mathbf{R}}_{An_1n_2}\big)
  \sin
  \Bigg(
       \frac{\pi\big(h+X_{An_2}\big)}{2h}
  \Bigg)~
  e^{i(K+q_N)Y_{n_1+\frac{n_2}{2}}}+
  \nonumber \\ &+&
  \frac{1}{\sqrt{N}}
  \sum_{n_1n_2}
  \Phi\big({\mathbf{r}}-{\mathbf{R}}_{Bn_1n_2}\big)
  \sin
  \Bigg(
       \frac{\pi\big(h+X_{Bn_2}\big)}{2h}
  \Bigg)~
  e^{i(K+q_N)Y_{n_1+\frac{n_2}{2}}},
  \label{wave-fun-CNT-QD}
\end{eqnarray}
where $\Phi({\mathbf{r}})$ is a Wannier function,
\begin{eqnarray}
  K &=&
  \frac{4\pi}{3a_0},
  \ \ \ \ \
  q_N ~=~
  \frac{2 \pi \mu}{3 N a_0},
  \label{qN-def}
\end{eqnarray}
where $N=3M+\mu$ with $M$ being integer and $\mu=\pm1$. $\mu$ is
chosen in such a way that $(K+q_N)Na_0$ be integer multiplier of
$2\pi$.

The Wannier functions $\Phi({\mathbf{r}}-{\mathbf{R}})$ and
$\Phi({\mathbf{r}}-{\mathbf{R}}')$ with
${\mathbf{R}}\neq{\mathbf{R}}'$ are orthogonal one to another,
so that the wave function (\ref{wave-fun-CNT-QD}) is normalized
by the condition,
$$
  \int{d}^{3}{\mathbf{r}}~
  \big|
      \psi_{\mathrm{CNT}}({\mathbf{r}})
  \big|^2
  ~=~ 1.
$$
The wave function $\psi_{\mathrm{CNT}}({\mathbf{r}})$ vanishes
when $|x|>h$. The term $q_N$ is introduced to satisfy the
Bohr-Sommerfeld quantization rule which says that $(K+q_N)Na_0$ is
an integer multiplier of $2\pi$.

In what following, we consider the following model wave function,
\begin{eqnarray}
  \psi_{\mathrm{CNT}}({\mathbf{r}}) &=&
  \frac{{\cal{N}}_{\mathrm{CNT}}}{\sqrt{\pi h}}~
  \sin
  \bigg(
      \frac{\pi(h+x)}{2h}
  \bigg)~
  e^{i (Kr_0+\nu) \phi}~
  e^{-\frac{|r-r_0|}{a_{\mathrm{C}}}},
  \label{wave-fun-CNT-QD-model}
\end{eqnarray}
where ${\mathbf{r}}=(x,r,\phi)$ are cylindrical coordinates,
$\nu=\frac{\mu}{3}$ [see Eq. (\ref{qN-def})],  $a_C=0.80$~{\AA}
is the radius of the carbon atom. The normalization
factor ${\cal{N}}_{\mathrm{CNT}}$ is
$$
  {\cal{N}}_{\mathrm{CNT}} ~=~
  \frac{2}{\sqrt{a_{\mathrm{C}}(4r_0+a_Ce^{-\frac{2r_0}{a_C}})}}
  ~\approx~
  \frac{1}{\sqrt{a_{\mathrm{C}} r_0}}.
$$

As for $\psi_6({\mathbf{r}})$, the wave function of the
6s-electron in the gold atom, we will use the following model wave function,
\begin{eqnarray}
  \psi_{6}({\mathbf{r}}) &=&
  \frac{1}{\sqrt{\pi a_{\mathrm{Au}}^{3}}}~
  e^{-\frac{\sqrt{x^2+r^2}}{a_{\mathrm{Au}}}},
  \label{wave-fun-Au-QD-model}
\end{eqnarray}
where $a_{\mathrm{Au}}=1.35$~{\AA} is the atomic radius of gold.

Substituting the wave function (\ref{wave-fun-CNT-QD-model}) into
eq. (\ref{V-exch-append}), we get
\begin{eqnarray}
  V_{\mathrm{exch}} &=&
  -\frac{2}{\pi h a_{\mathrm{Au}}^{3}a_{\mathrm{C}}r_0}
  \int\limits_{-h}^{h}dx~dx'~
  \int\limits_{0}^{\infty}rdr~r'dr'~
  {\cal{G}}(x,x',r,r')
  {\cal{F}}(x,x',r,r'),
  \label{V-exch-integral}
\end{eqnarray}
where
\begin{eqnarray}
  {\cal{G}}(x,x',r,r') &=&
  \sin
  \bigg(
       \frac{\pi(h+x)}{2h}
  \bigg)
  \sin
  \bigg(
       \frac{\pi(h+x')}{2h}
  \bigg)
  e^{-\frac{|r-r_0|+|r'-r_0|}{a_{\mathrm{C}}}}~
  e^{-\frac{1}{a_{\mathrm{Au}}}
     \big(
         \sqrt{x^2+r^2}+
         \sqrt{{x'}^2+{r'}^2}
     \big)},
  \label{G-def}
  \\
  {\cal{F}}(x,x',r,r') &=&
  \int\limits_{0}^{2\pi}d\phi~d\phi'~
  \frac{e^{-i (Kr_0+\nu) (\phi-\phi')}}
       {\sqrt{(x-x')^2+r^2+{r'}^2-2rr'\cos(\phi-\phi')}}.
  \label{F-elliptic-def}
\end{eqnarray}

\begin{figure}[htb]
\centering
 \includegraphics[width=55 mm,angle=0]
    {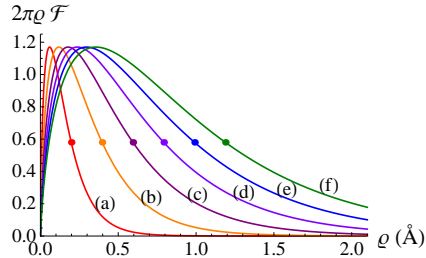}
 \caption{\footnotesize
  (Color online)
  The function $2\pi\varrho{\cal{F}}(x+\frac{1}{2}\varrho\cos\varphi,%
  x-\frac{1}{2}\varrho\cos\varphi,%
  r+\frac{1}{2}\varrho\sin\varphi,%
  r-\frac{1}{2}\varrho\sin\varphi)$
  as functions of $\varrho$ for different values of $r$.
  Curves (a) -- (f) correspond to
  $r=0.5,~1,~1.5,~2,~2.5,~3$~{\AA}.
  The dots denote
  the half-maximum of the function.}
 \label{Fig-F-elliptic}
\end{figure}

The function ${\cal{F}}(x,x',r,r')$ depends on $x-x'$ but not on
$x+x'$. It has its maximum at $x-x'=0$ and $r-r'=0$, decreases
with $\varrho=\sqrt{(x-x')^2+(r-r')^2}$ and vanishes
when $\varrho\gg\varrho_c$ exceeds some critical value
$\varrho_c$. Numerical calculations plotted in
Fig. \ref{Fig-F-elliptic} show that $\varrho_c\lesssim1.2$~{\AA}.
The function ${\cal{G}}(x,x',r,r')$ varies slowly with $x-x'$ and $r-r'$.
Therefore, we can approximate ${\cal{F}}(x,x',r,r')$ by
the following expression,
\begin{eqnarray}
  {\cal{F}}(x,x',r,r') &=&
  {\cal{F}}_0\big(r\big)
  \delta(r-r')
  \delta(x-x'),
\end{eqnarray}
where
\begin{eqnarray}
  {\cal{F}}_0(r) &=&
  \int\limits_{-\infty}^{\infty}dx'~
  \int\limits_{-\infty}^{\infty}dr'~
  {\cal{F}}
  \bigg(
       x+\frac{x'}{2},
       x-\frac{x'}{2},
       r+\frac{r'}{2},
       r-\frac{r'}{2}
  \bigg).
  \label{F0-def}
\end{eqnarray}

Integration of the RHS of Eq. (\ref{F0-def}) yields,
\begin{eqnarray*}
  {\cal{F}}_0(r) &=&
  \frac{32\pi^2r}{99}.
\end{eqnarray*}

Then the exchange interaction (\ref{V-exch-integral}) can be
written as,
\begin{eqnarray}
  V_{\mathrm{exch}} &=&
  -\frac{64\pi}
       {99 h a_{\mathrm{Au}}^{3} a_{\mathrm{C}} r_0}
  \int\limits_{-h}^{h}dx~
  \sin^2
  \bigg(
       \frac{\pi(h+x)}{2h}
  \bigg)
  \int\limits_{0}^{\infty}r^3dr~
  e^{-\frac{2|r-r_0|}{a_{\mathrm{C}}}}~
  e^{-\frac{2}{a_{\mathrm{Au}}}\sqrt{x^2+r^2}}.
  \label{V-exch-integral-approx}
\end{eqnarray}

For $a_{\mathrm{C}}=0.8$~{\AA}, $a_{\mathrm{Au}}=1.35$~{\AA},
$r_0=3.13$~{\AA} and $h=5r_0=15.65$~{\AA},
$V_{\mathrm{exch}}\approx-43.3$~meV. The exchange interaction
is ferromagnetic which agrees with the Hund rules.

\subsection{IV. Absence of electron tunneling between the CNT and
            the Caged Atom}

The tunneling rate $T$ between the atom and the CNT can be
estimated as,
\begin{eqnarray}
  V_t &=&
  \int{d^3{\mathbf{R}}}~
  \big|
      \Psi_{000}^{\mathrm{Au}}({\mathbf{R}})
  \big|^2
  \int{d^3{\mathbf{r}}}~
  \Psi_{\mathrm{CNT}}^{*}({\mathbf{r}})~
  \frac{e^2}{\big|{\mathbf{r}}+{\mathbf{R}}\big|}~
  \psi_6({\mathbf{r}}),
  \label{tunneling-atom-CNT-def}
\end{eqnarray}
where the electronic wave functions
$\Psi_{\mathrm{CNT}}({\mathbf{r}})$ and
$\psi_6({\mathbf{r}})$ are given by
Eqs. (\ref{wave-fun-CNT-QD-model}) and
(\ref{wave-fun-Au-QD-model}), and
the atomic wave function
$\Psi_{000}^{\mathrm{Au}}({\mathbf{R}})$
is defined by Eq. (\ref{subeqs-wave-fun-slow}).
Using cylindrical coordinates ${\mathbf{r}}=(x,r,\phi)$
and ${\mathbf{R}}=(X,R,\varphi)$, we get
\begin{eqnarray*}
  V_t &=&
  \frac{1}{\sqrt{\pi a_{\mathrm{Au}}^3ha_{\mathrm{C}}r_0}}~
  \int\limits_{-h}^{h}dx~
  \sin
  \bigg(
       \frac{\pi(h+x)}{2h}
  \bigg)
  \int\limits_{-\infty}^{\infty}dX~
  F_{0}^{2}(X)
  \int\limits_{0}^{\infty}rdr~
  e^{-\frac{1}{a_{\mathrm{C}}}|r-r_0|}
  e^{-\frac{1}{a_{\mathrm{Au}}}\sqrt{x^2+r^2}}
  \int\limits_{0}^{\infty}RdR~
  \Phi_{00}^{2}(R)
  \nonumber \\ && \times
  \int\limits_{0}^{2\pi}d\varphi~
  \frac{e^2}{\sqrt{\big(x-X\big)^2+r^2+R^2+2rR\cos\varphi}}~
  \int\limits_{0}^{2\pi}d\phi~
  e^{i(Kr_0+\nu)\phi}.
\end{eqnarray*}
Taking into account that $Kr_0+\nu$ is non-zero integer, we get
$V_t=0$.

\subsection{V. Derivation of the Scaling Equations}

In this subsection we describe the derivation of the scaling equations
as displayed in the main text around Eqs. (8-12) therein.
In order to carry out the poor man's scaling analysis, let us divide
the energy interval $|\epsilon|<D$ (``conduction band'')
into three intervals (see Fig. \ref{Fig-scaling}):
\begin{itemize}
\item[i.] $-D+\delta{D}<\epsilon-\epsilon_F<D-\delta{D}$,
\item[ii.] $D-\delta{D}<\epsilon-\epsilon_F<D$,
\item[iii.] $-D<\epsilon-\epsilon_F<-D+\delta{D}$.
\end{itemize}
The quantum states within the interval (i) are retained
and the quantum states within the intervals (ii) and (iii)
are to be integrated out.

\begin{figure}[htb]
\centering
 \begin{tabular}{|c|c|}
 \hline
 (a) & (b)
 \\
 ~
 \includegraphics[height = 45 mm,angle=0]
    {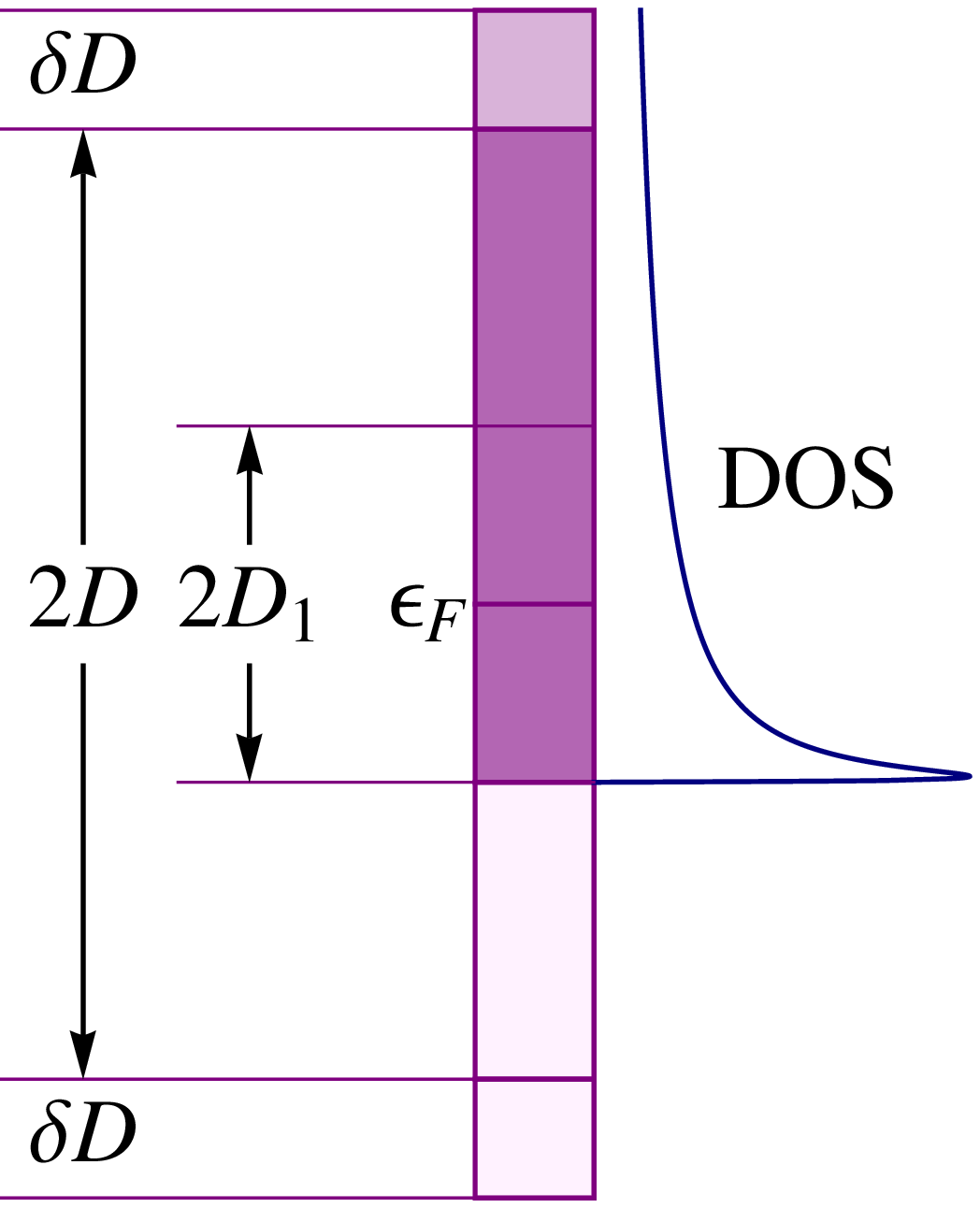}
 ~
 &
 ~
 \includegraphics[height = 45 mm,angle=0]
    {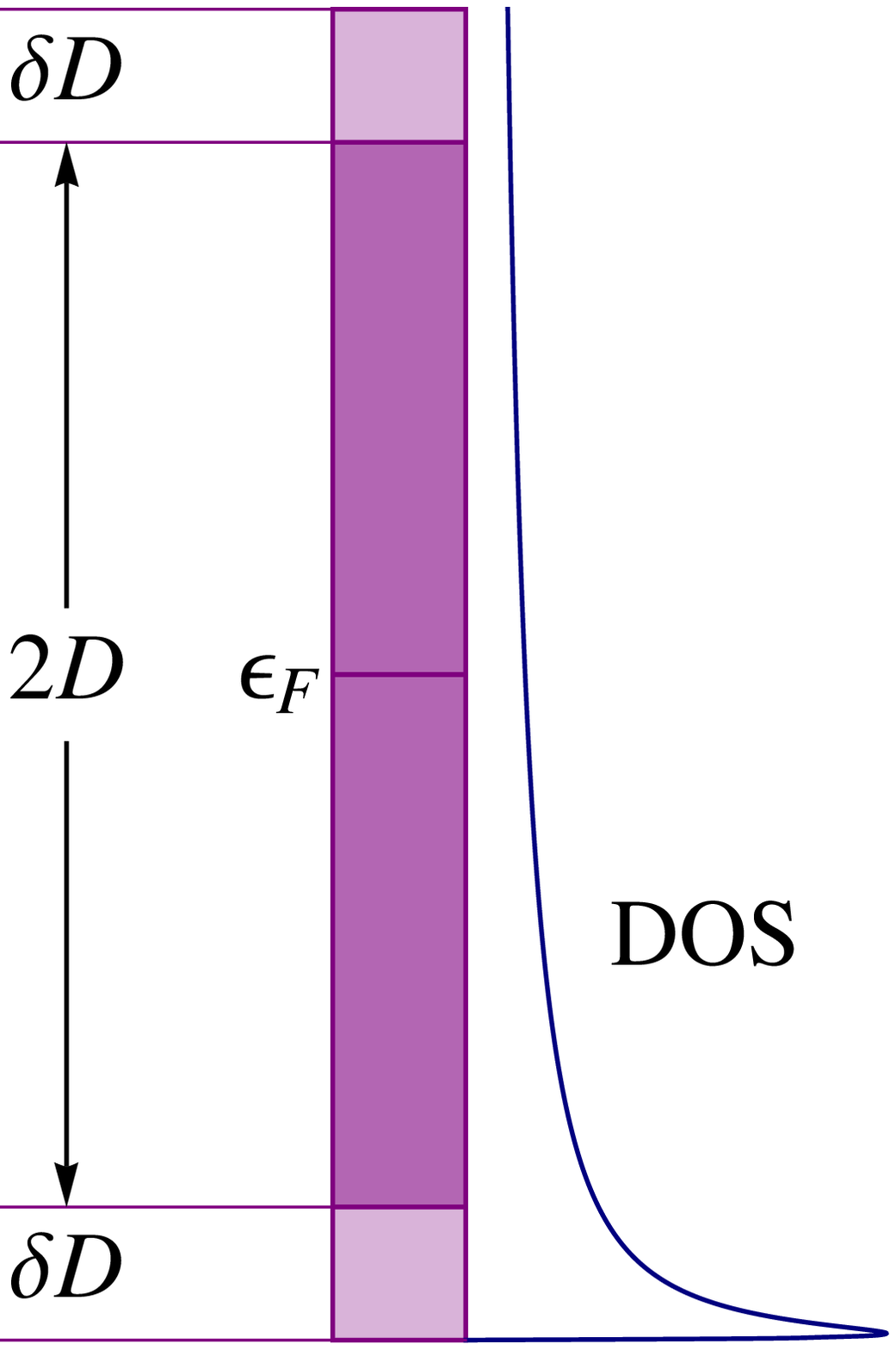}
 ~
 \\
 \hline
 \end{tabular}
 \caption{\footnotesize
  (Color online)
  The particle and hole states which are integrated out from
  the conduction band on reducing the bandwidth
  by $\delta{D}$ for $D>D_1$ [panel (a)] and
  $D<D_1$ [panel (b)]. The curve denotes the density
  of states (DOS).}
 \label{Fig-scaling}
\end{figure}

The corrections to $H_K$ due to the virtual scattering are
shown in Figs. \ref{Fig-diagrams-2nd} and \ref{Fig-diagrams-3rd}.
Here the solid blue lines
describe the quantum state of the quantum dot. The
dashed purple lines with a red circle at one side describe
falling or scattered conduction electron and dashed purple lines
with two red circles at the ends describe virtual conduction
electron within the energy interval (i). The dashed and dotted
violet lines with arrow right or left correspond to
virtual electron with energy within the interval (ii) or (iii).

\begin{figure}[htb]
\centering
 \begin{tabular}{|c|c|}
 \hline
 (a) & (b)
 \\
 ~
 \includegraphics[height=26 mm,angle=0]
    {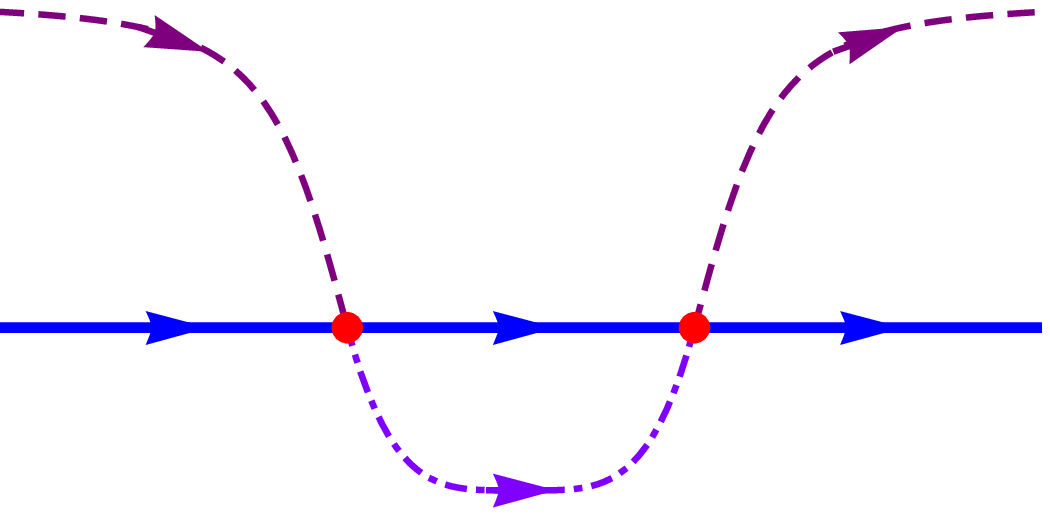}
 ~
 &
 ~
 \includegraphics[height=26 mm,angle=0]
    {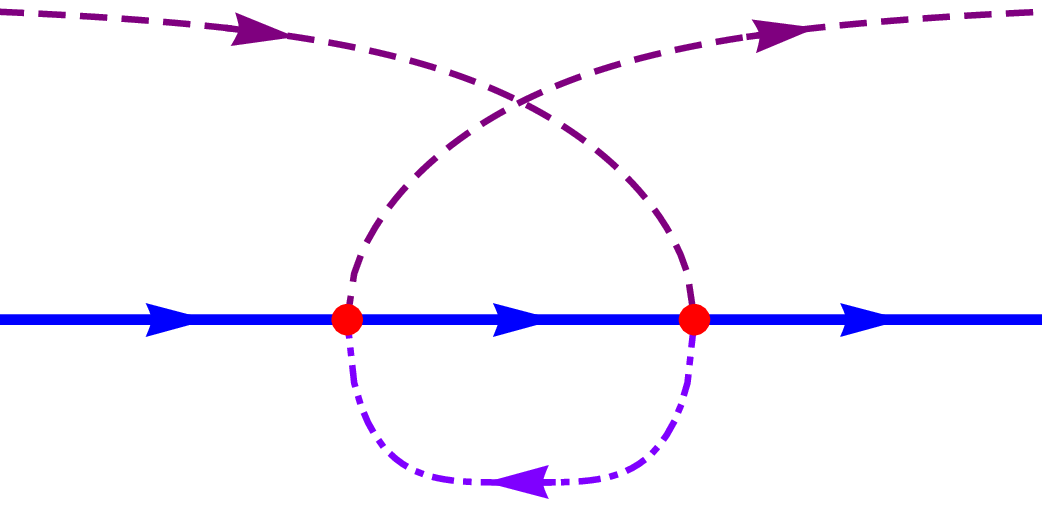}
 ~
 \\
 \hline
 \end{tabular}
 \caption{\footnotesize
  (Color online)
  Second order electronic [panel (a)] and hole [panel (b)]
  diagrams which have a particle in an intermediate state
  at a band edge (dashed and dotted line).}
 \label{Fig-diagrams-2nd}
\end{figure}

\begin{subequations}
The contribution of the second order diagram in
Fig. \ref{Fig-diagrams-2nd}(a) is,
\begin{eqnarray}
  \delta{K}^{(2a)}_{\epsilon_{q} \epsilon_{q'}}
  &=&
  -\frac{\delta{D}}{D}~
  \Big(
      K_{\epsilon_q D}
      K_{D \epsilon_{q'}}+
      \frac{3}{16}~
      J_{\epsilon_q D}
      J_{D \epsilon_{q'}}
  \Big)~
  \rho(D),
  \label{dK2a}
  \\
  \delta{J}^{(2a)}_{\epsilon_{q} \epsilon_{q'}}
  &=&
  -\frac{\delta{D}}{D}~
  \Big(
      K_{\epsilon_q D}
      J_{D \epsilon_{q'}}+
      J_{\epsilon_q D}
      K_{D \epsilon_{q'}}+
      \frac{1}{2}~
      J_{\epsilon_q D}
      J_{D \epsilon_{q'}}
  \Big)~
  \rho(D).
  \label{dJ2a}
\end{eqnarray}

\begin{figure}[htb]
\centering
 \begin{tabular}{|c|c|}
 \hline
 (a) & (b)
 \\
 ~
 \includegraphics[height=26 mm,angle=0]
    {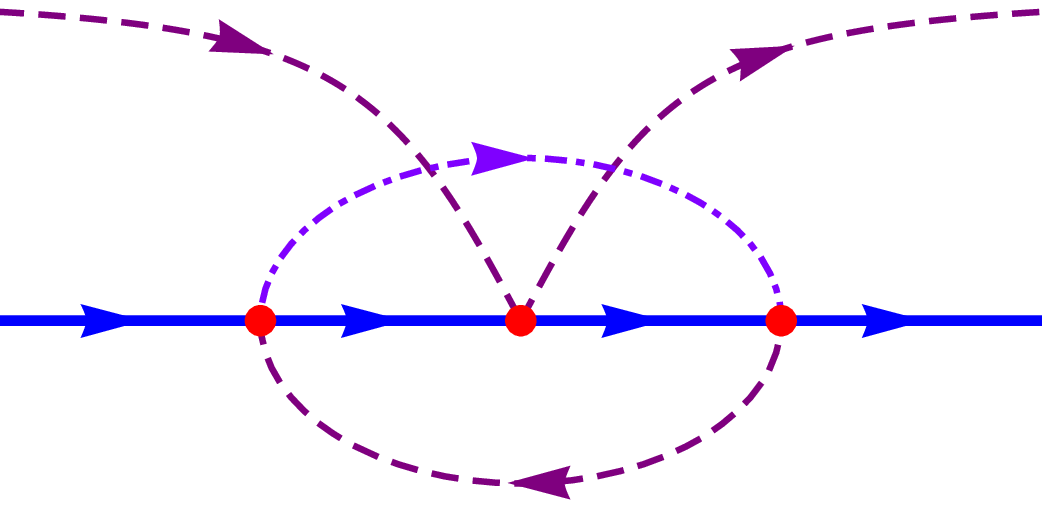}
 ~
 &
 ~
 \includegraphics[height=26 mm,angle=0]
    {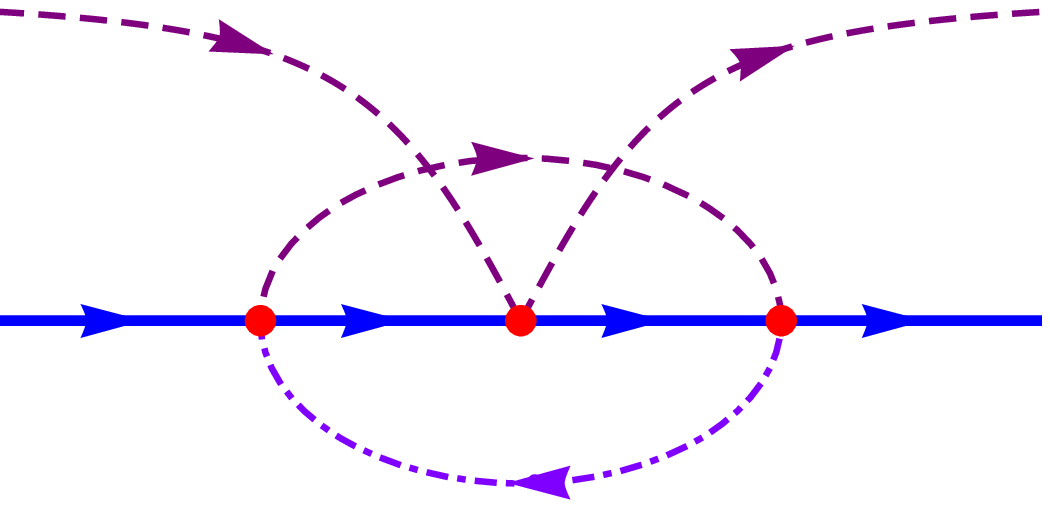}
 ~
 \\
 \hline
 \end{tabular}
 \caption{\footnotesize
  (Color online)
  Third order electronic [panel (a)] and hole [panel (b)]
  diagram with a particle in an intermediate state
  at a band edge (dashed and dotted line) which
  results in the over-screened fixed point.}
 \label{Fig-diagrams-3rd}
\end{figure}

The contribution of the second order diagram in
Fig. \ref{Fig-diagrams-2nd}(b) is,
\begin{eqnarray}
  \delta{K}^{(2b)}_{\epsilon_{q} \epsilon_{q'}}
  &=&
  \frac{\delta{D}}{D}~
  \Big(
      K_{\epsilon_q ~-D}
      K_{-D \epsilon_{q'}}+
      \frac{3}{16}~
      J_{\epsilon_q ~-D}
      J_{-D \epsilon_{q'}}
  \Big)~
  \rho(-D),
  \label{dK2b}
  \\
  \delta{J}^{(2b)}_{\epsilon_{q} \epsilon_{q'}}
  &=&
  \frac{\delta{D}}{D}~
  \Big(
      K_{\epsilon_q~-D}
      J_{-D \epsilon_{q'}}+
      J_{\epsilon_q~ -D}
      K_{-D \epsilon_{q'}}-
      \frac{1}{2}~
      J_{\epsilon_q D}
      J_{D \epsilon_{q'}}
  \Big)~
  \rho(-D).
  \label{dJ2b}
\end{eqnarray}
  \label{subeqs-d2}
\end{subequations}
It should be taken into account that $\rho(-D)=0$
when $D>D_1$, so that the diagram in Fig. \ref{Fig-diagrams-2nd}(b)
contributes to $H_K$ just when $D<D_1$.

The contribution of the third order diagram in
Fig. \ref{Fig-diagrams-3rd}(a) is,
\begin{eqnarray*}
  \delta{K}^{(3a)}_{\epsilon_{q} \epsilon_{q'}}
  &=&
  -\frac{\delta{D}}{D}~
  K_{\epsilon_q \epsilon_{q'}}
  \Big(
      8k^2+
      \frac{3}{2}~
      j^2
  \Big)~
  \frac{1}{D}
  \int\limits_{-D}^{0}d\epsilon~
  \vartheta(-\epsilon)
  \vartheta(\epsilon+D_1),
  \\
  \delta{J}^{(3a)}_{\epsilon_{q} \epsilon_{q'}}
  &=&
  \frac{\delta{D}}{D}~
  J_{\epsilon_q \epsilon_{q'}}
  \Big(
      -4k^2+
      j^2
  \Big)~
  \frac{1}{D}
  \int\limits_{-D}^{0}d\epsilon~
  \vartheta(-\epsilon)
  \vartheta(\epsilon+D_1).
\end{eqnarray*}

\begin{subequations}
After integration over $\epsilon$, we get,
\begin{eqnarray}
  \delta{K}^{(3a)}_{\epsilon_{q} \epsilon_{q'}}
  &=&
  -\frac{\delta{D}}{D}~
  K_{\epsilon_q \epsilon_{q'}}
  \Big(
      8k^2+
      \frac{3}{2}~
      j^2
  \Big)~
  \frac{{\min}(D,D_1)}{D},
  \label{dK3a}
  \\
  \delta{J}^{(3a)}_{\epsilon_{q} \epsilon_{q'}}
  &=&
  \frac{\delta{D}}{D}~
  K_{\epsilon_q \epsilon_{q'}}
  \Big(
      -4k^2+
      j^2
  \Big)~
  \frac{{\min}(D,D_1)}{D}.
  \label{dJ3a}
\end{eqnarray}

Similarly, the contribution of the third order diagram in
Fig. \ref{Fig-diagrams-3rd}(b) is,
\begin{eqnarray}
  \delta{K}^{(3b)}_{\epsilon_{q} \epsilon_{q'}}
  &=&
  \frac{\delta{D}}{D}~
  K_{\epsilon_q \epsilon_{q'}}
  \Big(
      8k^2+
      \frac{3}{2}~
      j^2
  \Big)~
  \vartheta(D_1-D),
  \label{dK3b}
  \\
  \delta{J}^{(3b)}_{\epsilon_{q} \epsilon_{q'}}
  &=&
  \frac{\delta{D}}{D}~
  K_{\epsilon_q \epsilon_{q'}}
  \Big(
      4k^2+
      j^2
  \Big)~
  \vartheta(D_1-D).
  \label{dJ3b}
\end{eqnarray}
  \label{subeqs-d3}
\end{subequations}

Combining Eqs. (\ref{subeqs-d2}) and (\ref{subeqs-d3}),
we get the scaling equations for the dimensionless
couplings $k$ and $j$. For ${D}\gg{D}_{1}$, the equations are
\begin{subequations}
\begin{eqnarray}
  \delta{k} &=&
  -\frac{\delta{D}}{D}~
  \Big(
      k^2+
      \frac{3j^2}{16}
  \Big),
  \label{dk-D>S1}
  \\
  \delta{j} &=&
  -\frac{\delta{D}}{D}~
  \Big(
      2kj+
      \frac{j^2}{2}
  \Big).
  \label{dj-D>d1}
\end{eqnarray}
  \label{subeqs-d-D>D1}
\end{subequations}
Approximating
$$
  D~\frac{\delta{k}}{\delta{D}} ~\approx~
  \frac{\partial{k}}{\partial\ln{D}},
  \ \ \ \ \
  D~\frac{\delta{j}}{\delta{D}} ~\approx~
  \frac{\partial{j}}{\partial\ln{D}},
$$
we get Eqs. (8a) and (8b) of the main text.
Similarly, for $D<D_1$ we get $\delta{k}=0$ and
\begin{eqnarray}
  \delta{j} &=&
  -\frac{\delta{D}}{D}~
  \Big(
      j^2-2j^3
  \Big).
  \label{dj-D<D1}
\end{eqnarray}
The last equation yields Eq. (10) of the main text.

\end{document}